\documentclass[12pt]{article}

\usepackage{color}
\usepackage{soul}

\usepackage{preamble}

\title{A Latent Variable Model for Response Times with Individual-Specific Change-Points}
\author{
Gabriel Wallin\\
School of Mathematical Sciences, Lancaster University
\and
Nivedita Bhaktha\\
Department of Management Sciences, Indian Institute of Technology Kanpur
}
\date{}

\begin{document}

\maketitle

\begin{abstract}
Response times collected in computerised assessments provide information about the underlying response process and may exhibit within-person variation over the course of a test. We propose a latent variable model for log response times that incorporates individual-specific change-points. The model extends the log-normal response time model by allowing an item-specific shift in the mean structure after an unobserved change-point. The change-point is treated as a discrete latent variable, and its distribution is modeled as a function of latent speed. Estimation is carried out using marginal maximum likelihood. The framework yields posterior distributions for change-point locations, allowing uncertainty to be quantified at the individual level, and supports statistical inference for the change-point effect parameters. A simulation study examines parameter recovery and change-point estimation under varying boundary conditions, prevalence of changers, sample sizes, and test lengths. The results show accurate recovery of item and structural parameters. The proposed model provides a unified approach to modeling response times with within-person changes in behaviour.
\end{abstract}

\section{Introduction}

Response times (RTs) are increasingly used as process data for studying how individuals interact with items, tasks, or survey questions. In addition to item responses, RTs reflect latent characteristics such as working speed, effort, and engagement, and have become an important component in psychometric analysis \citep{van2008using, lee2014modeling}. Because RTs are captured automatically during the response process, they can reveal behavioural patterns and processes that are not directly observable through response accuracy \citep{fan2012does}.

RT data have been shown to provide useful information about the latent traits that measurement instruments are designed to assess, especially when modeled jointly with item responses. In this paper, however, our focus is different: we use RTs to study within-person variation in response behaviour across a sequence of items or tasks. Such variation may arise from factors such as changes in effort, fatigue, shifts in response strategies, or time constraints, and may manifest as systematic changes in response times. These changes can affect the estimation of item parameters and latent traits, and may compromise the validity and interpretability of the inferences drawn from the instrument. Among these, test speededness has received considerable attention \citep{schnipke1997modeling, van1999using}. Speededness occurs when time constraints influence the response, often resulting in faster responses towards the end of a test and, in some cases, reduced response accuracy \citep{wise2005response, goegebeur2010person}. Rapid guessing and aberrant behaviour have also been extensively studied \citep{wise2006modeling, wise2010rapid, wise2012detecting, zhu2023bayesian, meade2012identifying, du2025detecting}. These behaviours can affect the estimation of item parameters and latent abilities, and may compromise the validity and fairness of test scores.

Several approaches have been proposed to model and detect such variation using RT data. Parametric models, such as the log-normal response time model \citep{van2006lognormal}, provide a framework for modeling RTs through item-specific time intensity parameters and person-specific speed parameters. Extensions of this framework include hierarchical and joint models for response accuracy and response times \citep{van2008using, vanderguo2008}. Mixture and latent class models distinguish between response strategies such as solution behaviour and rapid guessing \citep{wang2015mixture, marianti2014modeling}. Growth models allow response speed to vary over the course of a test \citep{fox2016joint}. While these approaches provide interpretable frameworks, they typically impose smooth or gradual changes in response behaviour.

An alternative approach treats the modeling of structural changes in response behaviour as a change-point problem. In this framework, the objective is to identify a point in the response sequence at which behaviour changes. Classical methods based on likelihood ratio, Wald, or score tests have been adapted to this setting \citep{sinharay2016person, sinharay2017detecting, cheng2022application, shao2016detecting}. Cumulative sum (CUSUM)-type procedures are commonly used because of their sensitivity to shifts in mean behaviour \citep{page1954continuous, yu2022cusum}. These methods have been applied in both offline and online settings, including real-time monitoring in computerised assessments \citep[e.g.,][]{sinharay2016person}. Despite their effectiveness, many of these approaches focus on detection rather than modeling, or treat the change-point as a fixed feature rather than a latent variable. A limitation of existing methods is therefore that they often identify changes in response behaviour without modeling the change-points themselves as individual-specific latent variables, making it difficult to study how such changes vary across individuals and relate to latent speed and item characteristics. Standard RT models typically assume that the individual speed is constant throughout the measurement process, while change-point methods are often applied as separate procedures. This separation limits the ability to model the joint relationship between latent speed, item characteristics, and behavioural changes.

Recent work by \citet{wallin2025latent} addresses this limitation in the context of item response data by introducing a latent variable model with person-specific change-points. Their approach enables joint estimation of item parameters, latent traits, and change-point locations, and shows that accounting for behavioural changes can reduce bias in ability estimates under time pressure. However, this framework is formulated for item responses and does not explicitly model RTs. This leaves open the question of how individual-specific change-points can be modeled using RTs, which often provide a strong signal of changes in response behaviour.

In this paper, we propose a change-point model for RTs that embeds individual-specific change-points within a latent variable model for response speed. The model builds on the log-normal RT framework and allows for an additive shift in log response times after an unobserved change-point. The change-point is treated as a discrete latent variable, and its distribution depends on the respondent's latent speed through a discrete-time hazard function. This formulation allows for heterogeneity in both response speed and the probability of behavioural change. The model is estimated by marginal maximum likelihood and posterior probabilities for the change-point location are obtained for each respondent, giving a natural uncertainty quantification of the detected change-points. 

The simulation study examines parameter recovery and change-point estimation under different values of the boundary parameter, the proportion of changers, sample size, and test length. The framework is also illustrated using real data from chess players solving strategic problems.

The remainder of the paper is organized as follows. Section~\ref{sec:model} presents the model and inference framework. Section~\ref{sec:simulation} describes the simulation study and reports the results. Section~\ref{sec:empirical} presents the empirical application. Section~\ref{sec:discussion} concludes with a discussion of the findings, limitations, and possible extensions.

\section{Model and Inference}\label{sec:model}

\subsection{Model Specification}\label{sec:model_spec}

Let $T_{ij}$ denote the response time of respondent $i$ on item $j$, where $i = 1, \ldots, N$ and $j = 1, \ldots, J$. We model the natural logarithm of response times as
\begin{equation*}
\log(T_{ij}) = \beta_j - \alpha_j \xi_i + \gamma_j \mathbb{I}(j > \tau_i) + \varepsilon_{ij},
\end{equation*}
where $\beta_j$ is the time intensity parameter for item $j$, $\alpha_j$ is the speed discrimination parameter, $\xi_i$ is the latent speed of respondent $i$, $\gamma_j$ is the change-point effect associated with item $j$, $\tau_i$ is the change-point location for respondent $i$, $\mathbb{I}(j > \tau_i)$ is an indicator function equal to 1 if item $j$ is beyond respondent  $i$'s change-point, and $\varepsilon_{ij} \sim N(0, \sigma_j^2)$ independently across respondents and items.

The latent speed variables are assumed to be independent and identically distributed as $\xi_i \sim N(0,1)$. Larger values of $\xi_i$ correspond to faster responding, as they reduce the expected log response time. The change-point $\tau_i$ takes values in the set $\{c+1, \ldots, J\}$, where $c < J$ is a fixed integer. For items $j \leq c+1$, we impose $\gamma_j = 0$ to ensure that the model is identifiable and that all respondents share a common baseline segment of at least $c+1$ items. The change-point $\tau_i$ is interpreted as the last item for which respondent $i$ follows the baseline response time process. For items $j > \tau_i$, the expected log response time is shifted by $\gamma_j$.

The parameters $\gamma_j$ characterise the magnitude and direction of the change in response behaviour after the change-point. Negative values of $\gamma_j$ correspond to shorter response times after the change-point and therefore indicate acceleration in responding, and positive values correspond to longer response times after the change-point. 

Conditional on $(\xi_i, \tau_i)$, the log response times $\{\log(T_{ij})\}_{j=1}^J$ are assumed to be independent with
\begin{equation*}
\log(T_{ij}) \mid \xi_i, \tau_i \sim N\left(\beta_j - \alpha_j \xi_i + \gamma_j \mathbb{I}(j > \tau_i), \sigma_j^2\right).
\end{equation*}
This specification thus extends the standard log-normal response time model by incorporating a structural change in the mean function at an individual-specific location. 

\subsection{Distribution of Change-Points}\label{sec:cpdist}

The change-point locations $\tau_i$ are modeled as discrete latent variables with support on $\{c+1, \ldots, J\}$. Their distribution is allowed to depend on the latent speed $\xi_i$, which induces dependence between the timing of behavioural changes and the baseline response process.

Let $\boldsymbol{\psi} = (\psi_1, \psi_2, \psi_3)^\top$ denote the parameters governing the distribution of $\tau_i$. Conditional on $\xi_i$, the probability mass function of $\tau_i$ is specified as
\begin{equation*}
P(\tau_i = k \mid \xi_i, \boldsymbol{\psi}) =
\begin{cases}
\displaystyle
\frac{\exp\{(k - c - 1)\psi_1\}}{\sum_{\ell=0}^{J-c-2} \exp(\ell \psi_1)}
\left[1 - \frac{1}{1 + \exp(-\psi_2 - \psi_3 \xi_i)}\right],
& k \in \{c+1, \ldots, J-1\}, \\[2ex]
\displaystyle
\frac{1}{1 + \exp(-\psi_2 - \psi_3 \xi_i)},
& k = J.
\end{cases}
\end{equation*}

The case $\tau_i = J$ corresponds to the absence of a change-point, in the sense that no structural shift occurs within the observed sequence. The probability of this event is governed by a logistic function of $\xi_i$, with intercept $\psi_2$ and slope $\psi_3$. This parameterisation allows the probability of exhibiting a change in behaviour to vary systematically with the latent speed.

Conditional on the event $\tau_i < J$, the distribution over admissible change-point locations is determined by the parameter $\psi_1$. This component defines a normalised exponential weighting over the indices $\{c+1, \ldots, J-1\}$, where the relative probability of later versus earlier change-points is controlled by the sign and magnitude of $\psi_1$. When $\psi_1 = 0$, the distribution over these locations is uniform. Positive values of $\psi_1$ assign higher probability to larger indices, while negative values favor earlier change-points.

This specification can be interpreted as a hazard-type model with an explicit no-change state: the logistic component controls the probability that no change occurs, while the weighting determines the location distribution among respondents who change. The dependence on $\xi_i$ connects the occurrence of behavioural change to the respondent's latent speed, which allows for heterogeneity in the change-point behaviour across respondents. 

\subsection{Marginal Likelihood and Parameter Estimation}\label{sec:estimation}

Let $\boldsymbol{y}_i = (y_{i1}, \ldots, y_{iJ})^\top$ denote the vector of log response times for respondent $i$, where $y_{ij}=\log(T_{ij})$, and let
$
\boldsymbol{\theta}
=
(\boldsymbol{\beta},\boldsymbol{\alpha},\boldsymbol{\gamma},
\boldsymbol{\sigma},\boldsymbol{\psi})
$
denote the full vector of model parameters. Conditional on the latent variables $(\xi_i,\tau_i)$, the density of $\boldsymbol{y}_i$ factorises as
\begin{equation*}
f(\boldsymbol{y}_i \mid \xi, \tau, \boldsymbol{\theta})
=
\prod_{j=1}^{J}
\frac{1}{\sqrt{2\pi\sigma_j^2}}
\exp\left\{
-\frac{1}{2\sigma_j^2}
\left[
y_{ij}
-
\beta_j
+
\alpha_j \xi
-
\gamma_j \mathbb{I}(j>\tau)
\right]^2
\right\}.
\end{equation*}

The full marginal likelihood is obtained by integrating over the latent speeds and summing over the possible change-point locations for all respondents:
\begin{equation*}
L(\boldsymbol{\theta})
=
\prod_{i=1}^{N}
\left[
\sum_{\tau=c+1}^{J}
\int
\left\{
\prod_{j=1}^{J}
\frac{1}{\sqrt{2\pi\sigma_j^2}}
\exp\left[
-\frac{
\left(
y_{ij}
-
\beta_j
+
\alpha_j \xi
-
\gamma_j \mathbb{I}(j>\tau)
\right)^2
}{
2\sigma_j^2
}
\right]
\right\}
P(\tau \mid \xi,\boldsymbol{\psi})
\phi(\xi)
\,d\xi
\right].
\end{equation*}

Since $P(\tau \mid \xi,\boldsymbol{\psi})$ depends on $\xi$, the summation over $\tau$ and integration over $\xi$ jointly marginalise over the discrete change-point location and the continuous latent speed. This dependence means that the uncertainty in the change-point location cannot be treated separately from the uncertainty in latent speed.

The one-dimensional integrals over $\xi$ are evaluated using Gaussian quadrature. With quadrature nodes $\{\xi^{(k)}\}_{k=1}^{K}$ and weights $\{w^{(k)}\}_{k=1}^{K}$, the likelihood is approximated by
\begin{equation*}
L(\boldsymbol{\theta})
\approx
\prod_{i=1}^{N}
\left[
\sum_{\tau=c+1}^{J}
\sum_{k=1}^{K}
w^{(k)}
f(\boldsymbol{y}_i \mid \xi^{(k)},\tau,\boldsymbol{\theta})
P(\tau \mid \xi^{(k)},\boldsymbol{\psi})
\right].
\end{equation*}
The corresponding marginal log-likelihood is maximised numerically using a quasi-Newton algorithm. The posterior weights and score equations used for implementation are provided in Appendix~\ref{app:score}.

A useful feature of the proposed formulation is that inference for the model parameters can be based directly on the marginal likelihood. This yields standard errors for the response-time parameters, including the change-point effects $\gamma_j$, as well as for the parameters $\boldsymbol{\psi}$ governing the distribution of change-point occurrence and timing. Under standard regularity conditions, the maximum likelihood estimator satisfies
\begin{equation*}
\sqrt{N}
\left(
\hat{\boldsymbol{\theta}}-\boldsymbol{\theta}_0
\right)
\xrightarrow{d}
N\left(
0,
\mathcal{I}(\boldsymbol{\theta}_0)^{-1}
\right),
\end{equation*}
where $\boldsymbol{\theta}_0$ denotes the true parameter value and
$\mathcal{I}(\boldsymbol{\theta}_0)$ is the Fisher information matrix for the observed data. In practice, we estimate the covariance matrix of $\hat{\boldsymbol{\theta}}$ using the inverse observed information matrix,
\begin{equation*}
\widehat{\mathrm{Var}}(\hat{\boldsymbol{\theta}})
=
\widehat{\mathcal{I}}(\hat{\boldsymbol{\theta}})^{-1},
\qquad
\widehat{\mathcal{I}}(\hat{\boldsymbol{\theta}})
=
-
\frac{\partial^2 \ell(\boldsymbol{\theta})}
{\partial \boldsymbol{\theta}\,\partial \boldsymbol{\theta}^{\top}}
\bigg|_{\boldsymbol{\theta}=\hat{\boldsymbol{\theta}}}.
\end{equation*}
Approximate Wald intervals for components of $\boldsymbol{\theta}$ are then obtained in the usual way. Thus, uncertainty about change-point behaviour is represented at two levels: respondent-specific uncertainty is summarised by the posterior distribution of $\tau_i$, while uncertainty about the population-level response-time and change-point distribution parameters is reflected in the standard errors for $\gamma_j$ and $\boldsymbol{\psi}$.

\subsection{Posterior Inference for Change-Points}\label{sec:postcp}

Given the maximum likelihood estimate $\hat{\boldsymbol{\theta}}$, inference on the change-point location for respondent $i$ is based on the posterior distribution of $\tau_i$ conditional on the observed response-time sequence $\boldsymbol{y}_i$. This distribution is obtained by marginalising over the latent speed variable $\xi_i$. For $\tau \in \{c+1,\ldots,J\}$,
\begin{equation*}
P(\tau_i=\tau \mid \boldsymbol{y}_i,\hat{\boldsymbol{\theta}})
=
\int
p_i(\tau,\xi \mid \boldsymbol{y}_i,\hat{\boldsymbol{\theta}})
\,d\xi,
\end{equation*}
where
\begin{equation*}
p_i(\tau,\xi \mid \boldsymbol{y}_i,\hat{\boldsymbol{\theta}})
=
\frac{
f(\boldsymbol{y}_i \mid \xi,\tau,\hat{\boldsymbol{\theta}})
P(\tau \mid \xi,\hat{\boldsymbol{\psi}})
\phi(\xi)
}{
\sum_{\tau'=c+1}^{J}
\int
f(\boldsymbol{y}_i \mid u,\tau',\hat{\boldsymbol{\theta}})
P(\tau' \mid u,\hat{\boldsymbol{\psi}})
\phi(u)
\,du
}.
\end{equation*}

A point estimate can be obtained as the posterior mode,
\begin{equation*}
\hat{\tau}_i
=
\arg\max_{\tau \in \{c+1,\ldots,J\}}
P(\tau_i=\tau \mid \boldsymbol{y}_i,\hat{\boldsymbol{\theta}}),
\end{equation*}
which corresponds to the most probable change-point location under the fitted model. The posterior distribution allows for direct quantification of uncertainty. For example, a plug-in posterior credible set for $\tau_i$ can be constructed by selecting a subset $\mathcal{T}_i \subset \{c+1,\ldots,J\}$ such that
\begin{equation*}
\sum_{\tau \in \mathcal{T}_i}
P(\tau_i=\tau \mid \boldsymbol{y}_i,\hat{\boldsymbol{\theta}})
\geq 1-\alpha,
\end{equation*}
for a specified level $\alpha \in (0,1)$. A natural choice is the smallest such subset, obtained by ordering the possible values of $\tau$ according to their posterior probabilities. This gives a highest-posterior-probability credible set for the respondent-specific change-point location.

The posterior probability of the event $\tau_i<J$ is
\begin{equation*}
P(\tau_i<J \mid \boldsymbol{y}_i,\hat{\boldsymbol{\theta}})
=
\sum_{\tau=c+1}^{J-1}
P(\tau_i=\tau \mid \boldsymbol{y}_i,\hat{\boldsymbol{\theta}}).
\end{equation*}
This quantity summarises the evidence that respondent $i$ experienced a structural change within the observed sequence. Conversely, $P(\tau_i=J \mid \boldsymbol{y}_i,\hat{\boldsymbol{\theta}})$ gives the posterior probability that no change-point occurred.

\subsection{Model Selection}\label{sec:model_selection}

The boundary parameter $c$ determines the earliest admissible change-point location and therefore controls the set of possible respondent-specific change-points. Smaller values of $c$ allow changes to occur earlier in the sequence of items and therefore provide a larger set of possible locations, whereas larger values of $c$ impose a longer segment modelled solely by the baseline model. In empirical applications, $c$ can be selected by fitting the model over a prespecified set of candidate values and comparing the resulting fitted models.

Let $\mathcal{C}$ denote a set of candidate values for $c$. For each $c\in\mathcal{C}$, the model is estimated by marginal maximum likelihood, yielding the maximised log-likelihood $\ell_c(\hat{\boldsymbol{\theta}}_c)$. Standard likelihood-based information criteria, such as AIC and BIC, may then be used to compare competing values of $c$ or alternative specifications of the change-point distribution. For example,
\[
\mathrm{BIC}(c)
=
-2\ell_c(\hat{\boldsymbol{\theta}}_c)
+
d_c\log N,
\]
where $d_c$ is the number of estimated parameters under the model with boundary parameter $c$. Because the model contains respondent-specific latent change-point locations, model selection may also be guided by criteria that account for classification uncertainty. In particular, the integrated completed likelihood criterion (ICL) augments the BIC with an entropy penalty based on the posterior distribution of the latent change-point locations. In the present setting, an ICL-type criterion can be written as
\[
\mathrm{ICL}(c)
=
\mathrm{BIC}(c)
+
2
\sum_{i=1}^{N}
\sum_{\tau=c+1}^{J}
-\hat{p}_{i\tau}
\log \hat{p}_{i\tau}.
\]
where
$
\hat{p}_{i\tau}
=
P(\tau_i=\tau\mid \boldsymbol{y}_i,\hat{\boldsymbol{\theta}}_c)
$
is the plug-in posterior probability of change-point location $\tau$ for respondent $i$ under the fitted model. Since $\sum_{\tau}\hat{p}_{i\tau}\log\hat{p}_{i\tau}\leq 0$, this criterion penalises models with diffuse posterior classifications of the change-point locations. Equivalently, it favours models that achieve good fit while yielding well-separated and interpretable latent change-point assignments.

The choice of criterion should reflect the purpose of the analysis. AIC may be useful when predictive fit is the primary goal, BIC provides a stronger penalty for model complexity, and ICL is particularly relevant when the interpretation of respondent-specific change-point locations is important. 



\section{Simulation study}\label{sec:simulation}

\subsection{Design}

The simulation study evaluates the performance of the proposed estimator, considering two grids of conditions. The primary grid isolates the effects of the earliest
admissible change-point location and the prevalence of changers, holding
sample size and test length fixed at the values for the empirical
analysis ($N = 256$, $J = 20$; see Section \ref{sec:empirical}). The boundary parameter $c$ takes values
in $\{5, 10, 15\}$, placing the earliest admissible change-point at
items 6, 11, and 16 respectively. The proportion of changers $\pi$
takes values in $\{0.15, 0.25, 0.40\}$, which covers the empirically observed rate of approximately 15\% as well as a substantially higher rate
that would be common in other empirical settings. This yields $3 \times 3 = 9$ conditions.

The secondary grid keeps the change-point position and prevalence fixed at values matching the empirical analysis ($\pi = 0.15$, $c$ set to approximately 60\% through the test) while varying sample size $N \in \{200, 600, 1800\}$ and test length $J \in \{20, 30, 40\}$. For $J = 20$, $30$, and $40$, the boundary parameter is set to $c = 12$, $18$, and $24$ respectively, keeping the proportional position of the earliest admissible change-point constant across test lengths. The three levels of $N$ span a range from slightly below the empirical sample to a more large-scale scenario of 1800 respondents. This yields a further $3 \times 3 = 9$ conditions, making it 18 conditions in total. In both grids, $c$ is treated as known and fixed at its true value.

\subsection{Data generation}

For each replication and condition, latent speed scores are drawn as
$\xi_i \sim \mathcal{N}(0, 1)$ for $i = 1, \ldots, N$. Change-point
locations $\tau_i$ are drawn independently for each respondent from
the model-implied distribution
\begin{equation*}
P(\tau_i = t \mid \xi_i) =
\begin{cases}
\left(1-\pi_i^{(\mathrm{no})}\right) q_t,
& t \in \{c+1,\ldots,J-1\}, \\[1ex]
\pi_i^{(\mathrm{no})},
& t=J,
\end{cases}
\end{equation*}
where \[
\pi_i^{(\mathrm{no})}
=
\frac{1}{1+\exp(-\psi_2-\psi_3\xi_i)}
\]
is the respondent-specific probability of no change-point and 
\[
q_t
=
\frac{\exp\{(t-c-1)\psi_1\}}
{\sum_{\ell=0}^{J-c-2}\exp(\ell\psi_1)}.
\]
is the discrete distribution over change-point locations given that a
change occurs, parametrised by $\psi_1$ as described in
Section~\ref{sec:model}. Log response times are then generated as
\begin{equation*}
  \log T_{ij} \sim \mathcal{N}\!\left(
    \beta_j - \alpha_j \xi_i + \gamma_j \mathbf{1}(j > \tau_i),\;
    \sigma_j^2
  \right),
\end{equation*}
independently across respondents and items.

True item parameters are drawn from: $\beta_j \sim \mathcal{U}(3, 4)$,
$\alpha_j \sim \mathcal{U}(0.5, 1.5)$,
$\gamma_j \sim \mathcal{U}(0.3, 0.8)$ for $j > c + 1$ and
$\gamma_j = 0$ otherwise, and
$\sigma_j \sim \mathcal{U}(0.2, 0.4)$. The interval $\mathcal{U}(3,4)$ was chosen to place the baseline response times on a realistic logarithmic second scale, approximately corresponding to item-level response times between 20 and 55 seconds before accounting for respondent speed and change-point effects. We use positive values of $\gamma_j$ to generate a post-change slowing effect. The proposed model, however, does not constrain the sign of $\gamma_j$; negative values correspond to acceleration after the change-point. Thus, the empirical application in Section \ref{sec:empirical}, in which the estimated effects are negative, represents the same structural mechanism with the opposite substantive direction.

The structural parameters are set to
$\psi_1 = 0.2$ and $\psi_3 = -0.5$; the value of $\psi_2$ is set per
condition to yield the target prevalence $\pi$ at $\xi_i = 0$ via
$\psi_2 = \log(\pi^{(\text{no})} / (1 - \pi^{(\text{no})}))$ where
$\pi^{(\text{no})} = 1 - \pi$. The value $\psi_3 = -0.5$ implies a moderate negative association between latent speed and the probability of exhibiting a change-point: respondents one standard deviation above average speed have a lower probability of changing response style than those at average speed.

\subsection{Evaluation}

Parameter recovery is assessed via bias and root mean squared error (RMSE), computed across replications for each condition. Change-point recovery is assessed via the mean absolute error of the respondent-specific change-point estimate,
\begin{equation*}
  \text{MAE}(\hat{\tau}) =
  \frac{1}{N} \sum_{i=1}^{N} |\hat{\tau}_i - \tau_i|,
\end{equation*}
averaged across replications, where $\hat{\tau}_i$ is either the modal
or the posterior mean change-point estimate. Both are reported to assess
whether point or distributional summaries better recover the true
change-point locations.

\subsection{Simulation Study Results}

\subsubsection{Primary grid}

Table~\ref{tab:sim_primary_cp} reports change-point recovery and
structural parameter estimates for the primary grid. The conditions with the latest boundary parameter ($c = 15$, earliest change-point at item 16) yield the lowest MAE for both the modal and posterior mean
change-point estimate, with MAE(mode) ranging from 0.128 to 0.226
across the three prevalence levels. This is natural since a later
boundary parameter restricts the change-point to a narrower window
of only four items ($\tau_i \in \{16, 17, 18, 19, 20\}$), making
individual change-points easier to locate precisely. The $c = 5$
conditions, where the change-point can occur anywhere in a 15-item
window, are harder to resolve and yield higher MAE, particularly when
prevalence is high ($c = 5$, $\pi = 0.40$: MAE(mode) $= 0.758$).

The effect of prevalence on change-point recovery is not monotone
and depends on the window width. At $c = 5$, higher prevalence
worsens $\tau$ recovery because more respondents with change-points spreading across many items increase uncertainty. At $c = 10$ and $c = 15$ the relationship is weaker, with the $\pi = 0.25$ condition sometimes outperforming $\pi = 0.15$. This likely reflects the fact that more changers provide more information for estimating the structural parameters $\psi_1$ and
$\psi_3$.

Structural bias is small and close to zero across all conditions. The largest exception is $\psi_2$ under $c = 10$, $\pi = 0.15$, where RMSE reaches 0.749, notably higher than in
other conditions. This condition combines a moderately wide
change-point window with a small proportion of changers, making
the no-change probability harder to estimate precisely. The
corresponding RMSE for $\psi_3$ is also larger (0.551) in this condition.

\begin{table}[t]
\centering
\caption{Simulation results (primary grid): change-point recovery and structural parameters. $N = 256$, $J = 20$ fixed; rows vary $c \in \{5, 10, 15\}$ and $\pi \in \{0.15, 0.25, 0.40\}$.}
\label{tab:sim_primary_cp}
\small
\resizebox{\textwidth}{!}{%
\begin{tabular}{@{}lrr rr rr rr@{}} 
\toprule
Condition & \multicolumn{2}{c}{MAE($\hat{\tau}$)} & \multicolumn{2}{c}{$\psi_1$} & \multicolumn{2}{c}{$\psi_2$} & \multicolumn{2}{c}{$\psi_3$} \\
\cmidrule(lr){2-3}\cmidrule(lr){4-5} \cmidrule(lr){6-7}\cmidrule(lr){8-9}
 & Mode & Mean & Bias & RMSE & Bias & RMSE & Bias & RMSE \\ 
\midrule
$c=5$, $\pi=0.15$ & $0.450$ & $0.558$ & $-0.041$ & $0.146$ & $-0.169$ & $0.447$ & $\phantom{-}0.037$ & $0.349$ \\
$c=5$, $\pi=0.25$ & $0.466$ & $0.528$ & $-0.016$ & $0.054$ & $\phantom{-}0.033$ & $0.241$ & $-0.027$ & $0.198$ \\
$c=5$, $\pi=0.4$ & $0.758$ & $0.831$ & $-0.050$ & $0.237$ & $\phantom{-}0.016$ & $0.270$ & $\phantom{-}0.020$ & $0.252$ \\
$c=10$, $\pi=0.15$ & $0.558$ & $0.681$ & $-0.071$ & $0.247$ & $-0.265$ & $0.749$ & $\phantom{-}0.058$ & $0.551$ \\
$c=10$, $\pi=0.25$ & $0.307$ & $0.357$ & $-0.019$ & $0.071$ & $-0.003$ & $0.224$ & $\phantom{-}0.011$ & $0.201$ \\
$c=10$, $\pi=0.4$ & $0.625$ & $0.692$ & $-0.042$ & $0.089$ & $\phantom{-}0.079$ & $0.310$ & $-0.032$ & $0.291$ \\
$c=15$, $\pi=0.15$ & $0.128$ & $0.160$ & $-0.039$ & $0.214$ & $-0.032$ & $0.250$ & $-0.024$ & $0.245$ \\
$c=15$, $\pi=0.25$ & $0.210$ & $0.254$ & $-0.015$ & $0.192$ & $-0.019$ & $0.228$ & $-0.016$ & $0.199$ \\
$c=15$, $\pi=0.4$ & $0.226$ & $0.274$ & $-0.004$ & $0.140$ & $-0.005$ & $0.189$ & $-0.017$ & $0.183$ \\ 
\bottomrule
\end{tabular}}
\begin{tablenotes}\small
\item MAE($\hat{\tau}$): mean absolute error of the modal (Mode) and posterior mean (Mean) change-point estimate. True values: $\psi_1 = 0.2$, $\psi_3 = -0.5$; $\psi_2$ set per condition to match target prevalence $\pi$. 
\end{tablenotes}
\end{table}

Table~\ref{tab:sim_primary_beta} presents the bias and RMSE for the item time intensity parameters $\hat{\beta}_j$ across all items and
conditions. The bias is falling within $[-0.026, 0.020]$ and the large majority within $[-0.015, 0.015]$. There is no indication of systematic over- or underestimation of item time intensity. RMSE ranges from approximately 0.038 to 0.125, reflecting the expected sampling variability at $N = 256$. There is no apparent pattern in either bias or RMSE across items, conditions, or values of $c$ and $\pi$, confirming that $\hat{\beta}_j$ estimation is robust to variation in the change-point structure.

\begin{table}[t]
\centering
\caption{Simulation results (primary grid): item time intensity $\hat{\beta}_j$. $N = 256$, $J = 20$.}
\label{tab:sim_primary_beta}
\small
\resizebox{\textwidth}{!}{%
\begin{tabular}{@{}rrrrrrrrrrrrrrrrrrr@{}}
\toprule
Item & \multicolumn{2}{c}{$c=5$, $\pi=0.15$} & \multicolumn{2}{c}{$c=5$, $\pi=0.25$} & \multicolumn{2}{c}{$c=5$, $\pi=0.4$} & \multicolumn{2}{c}{$c=10$, $\pi=0.15$} & \multicolumn{2}{c}{$c=10$, $\pi=0.25$} & \multicolumn{2}{c}{$c=10$, $\pi=0.4$} & \multicolumn{2}{c}{$c=15$, $\pi=0.15$} & \multicolumn{2}{c}{$c=15$, $\pi=0.25$} & \multicolumn{2}{c}{$c=15$, $\pi=0.4$} \\
\cmidrule(lr){2-3}\cmidrule(lr){4-5}\cmidrule(lr){6-7}\cmidrule(lr){8-9}\cmidrule(lr){10-11}\cmidrule(lr){12-13}\cmidrule(lr){14-15}\cmidrule(lr){16-17}\cmidrule(lr){18-19}
 & Bias & RMSE & Bias & RMSE & Bias & RMSE & Bias & RMSE & Bias & RMSE & Bias & RMSE & Bias & RMSE & Bias & RMSE & Bias & RMSE \\
\midrule
1 & $-0.008$ & $0.074$ & $\phantom{-}0.004$ & $0.098$ & $\phantom{-}0.006$ & $0.066$ & $\phantom{-}0.011$ & $0.113$ & $\phantom{-}0.003$ & $0.051$ & $\phantom{-}0.001$ & $0.064$ & $-0.001$ & $0.058$ & $\phantom{-}0.003$ & $0.042$ & $\phantom{-}0.009$ & $0.070$ \\
2 & $-0.007$ & $0.087$ & $\phantom{-}0.000$ & $0.100$ & $\phantom{-}0.008$ & $0.081$ & $\phantom{-}0.012$ & $0.085$ & $\phantom{-}0.006$ & $0.093$ & $\phantom{-}0.003$ & $0.102$ & $-0.002$ & $0.087$ & $\phantom{-}0.007$ & $0.103$ & $\phantom{-}0.013$ & $0.062$ \\
3 & $-0.008$ & $0.049$ & $\phantom{-}0.001$ & $0.071$ & $\phantom{-}0.010$ & $0.082$ & $\phantom{-}0.011$ & $0.082$ & $\phantom{-}0.006$ & $0.069$ & $\phantom{-}0.003$ & $0.121$ & $\phantom{-}0.002$ & $0.085$ & $\phantom{-}0.007$ & $0.090$ & $\phantom{-}0.009$ & $0.048$ \\
4 & $-0.006$ & $0.068$ & $-0.001$ & $0.042$ & $\phantom{-}0.006$ & $0.060$ & $\phantom{-}0.014$ & $0.086$ & $\phantom{-}0.009$ & $0.105$ & $\phantom{-}0.003$ & $0.063$ & $\phantom{-}0.000$ & $0.091$ & $\phantom{-}0.005$ & $0.046$ & $\phantom{-}0.006$ & $0.057$ \\
5 & $-0.010$ & $0.083$ & $\phantom{-}0.000$ & $0.090$ & $\phantom{-}0.010$ & $0.090$ & $\phantom{-}0.005$ & $0.058$ & $\phantom{-}0.007$ & $0.071$ & $-0.001$ & $0.073$ & $-0.002$ & $0.087$ & $\phantom{-}0.008$ & $0.104$ & $\phantom{-}0.011$ & $0.070$ \\
6 & $-0.007$ & $0.070$ & $\phantom{-}0.001$ & $0.043$ & $\phantom{-}0.009$ & $0.088$ & $\phantom{-}0.008$ & $0.069$ & $\phantom{-}0.003$ & $0.045$ & $\phantom{-}0.003$ & $0.104$ & $\phantom{-}0.004$ & $0.117$ & $\phantom{-}0.005$ & $0.095$ & $\phantom{-}0.008$ & $0.052$ \\
7 & $-0.001$ & $0.038$ & $\phantom{-}0.002$ & $0.073$ & $\phantom{-}0.013$ & $0.098$ & $\phantom{-}0.005$ & $0.048$ & $\phantom{-}0.006$ & $0.098$ & $\phantom{-}0.003$ & $0.056$ & $\phantom{-}0.000$ & $0.125$ & $\phantom{-}0.002$ & $0.045$ & $\phantom{-}0.017$ & $0.109$ \\
8 & $-0.008$ & $0.056$ & $\phantom{-}0.001$ & $0.074$ & $\phantom{-}0.017$ & $0.107$ & $\phantom{-}0.010$ & $0.073$ & $\phantom{-}0.006$ & $0.082$ & $-0.001$ & $0.112$ & $\phantom{-}0.005$ & $0.118$ & $\phantom{-}0.008$ & $0.104$ & $\phantom{-}0.014$ & $0.076$ \\
9 & $-0.006$ & $0.057$ & $\phantom{-}0.002$ & $0.071$ & $\phantom{-}0.015$ & $0.101$ & $\phantom{-}0.010$ & $0.072$ & $\phantom{-}0.006$ & $0.081$ & $\phantom{-}0.005$ & $0.070$ & $\phantom{-}0.000$ & $0.075$ & $\phantom{-}0.002$ & $0.073$ & $\phantom{-}0.007$ & $0.066$ \\
10 & $-0.011$ & $0.065$ & $\phantom{-}0.002$ & $0.060$ & $\phantom{-}0.015$ & $0.098$ & $\phantom{-}0.011$ & $0.076$ & $\phantom{-}0.005$ & $0.111$ & $\phantom{-}0.001$ & $0.047$ & $-0.001$ & $0.117$ & $\phantom{-}0.008$ & $0.092$ & $\phantom{-}0.012$ & $0.087$ \\
11 & $-0.008$ & $0.056$ & $\phantom{-}0.004$ & $0.073$ & $\phantom{-}0.013$ & $0.110$ & $\phantom{-}0.010$ & $0.070$ & $\phantom{-}0.006$ & $0.075$ & $-0.000$ & $0.112$ & $-0.001$ & $0.061$ & $\phantom{-}0.003$ & $0.053$ & $\phantom{-}0.009$ & $0.060$ \\
12 & $-0.010$ & $0.050$ & $\phantom{-}0.004$ & $0.103$ & $\phantom{-}0.009$ & $0.083$ & $\phantom{-}0.009$ & $0.105$ & $\phantom{-}0.006$ & $0.071$ & $\phantom{-}0.002$ & $0.118$ & $\phantom{-}0.001$ & $0.114$ & $\phantom{-}0.004$ & $0.105$ & $\phantom{-}0.020$ & $0.115$ \\
13 & $-0.008$ & $0.049$ & $\phantom{-}0.002$ & $0.104$ & $\phantom{-}0.011$ & $0.112$ & $-0.003$ & $0.048$ & $-0.001$ & $0.044$ & $-0.003$ & $0.106$ & $-0.002$ & $0.097$ & $\phantom{-}0.009$ & $0.097$ & $\phantom{-}0.020$ & $0.120$ \\
14 & $-0.011$ & $0.043$ & $\phantom{-}0.000$ & $0.047$ & $\phantom{-}0.009$ & $0.078$ & $\phantom{-}0.005$ & $0.104$ & $\phantom{-}0.004$ & $0.057$ & $-0.001$ & $0.082$ & $-0.001$ & $0.049$ & $\phantom{-}0.001$ & $0.078$ & $\phantom{-}0.010$ & $0.044$ \\
15 & $-0.016$ & $0.066$ & $-0.001$ & $0.048$ & $\phantom{-}0.004$ & $0.057$ & $\phantom{-}0.007$ & $0.113$ & $\phantom{-}0.005$ & $0.059$ & $-0.002$ & $0.084$ & $-0.001$ & $0.088$ & $\phantom{-}0.001$ & $0.067$ & $\phantom{-}0.016$ & $0.102$ \\
16 & $-0.012$ & $0.085$ & $\phantom{-}0.004$ & $0.081$ & $\phantom{-}0.008$ & $0.082$ & $-0.004$ & $0.073$ & $\phantom{-}0.002$ & $0.068$ & $-0.001$ & $0.068$ & $\phantom{-}0.003$ & $0.122$ & $\phantom{-}0.002$ & $0.058$ & $\phantom{-}0.017$ & $0.102$ \\
17 & $-0.005$ & $0.048$ & $\phantom{-}0.000$ & $0.051$ & $\phantom{-}0.014$ & $0.093$ & $\phantom{-}0.002$ & $0.052$ & $\phantom{-}0.003$ & $0.061$ & $-0.001$ & $0.090$ & $\phantom{-}0.001$ & $0.080$ & $\phantom{-}0.004$ & $0.063$ & $\phantom{-}0.008$ & $0.064$ \\
18 & $-0.019$ & $0.077$ & $-0.003$ & $0.109$ & $\phantom{-}0.007$ & $0.089$ & $\phantom{-}0.005$ & $0.106$ & $\phantom{-}0.001$ & $0.049$ & $-0.001$ & $0.096$ & $-0.002$ & $0.117$ & $\phantom{-}0.005$ & $0.076$ & $\phantom{-}0.009$ & $0.066$ \\
19 & $-0.010$ & $0.044$ & $-0.001$ & $0.060$ & $\phantom{-}0.015$ & $0.113$ & $-0.001$ & $0.102$ & $\phantom{-}0.004$ & $0.062$ & $-0.002$ & $0.079$ & $-0.004$ & $0.106$ & $\phantom{-}0.000$ & $0.094$ & $\phantom{-}0.008$ & $0.073$ \\
20 & $-0.026$ & $0.095$ & $-0.005$ & $0.057$ & $\phantom{-}0.008$ & $0.076$ & $-0.008$ & $0.087$ & $-0.001$ & $0.089$ & $\phantom{-}0.016$ & $0.100$ & $-0.005$ & $0.049$ & $-0.000$ & $0.103$ & $\phantom{-}0.009$ & $0.091$ \\
\bottomrule
\end{tabular}}
\begin{tablenotes}\small
\item Rows correspond to items; columns to simulation conditions. Cells show bias and RMSE for $\beta_j$.
\end{tablenotes}
\end{table}

Table~\ref{tab:sim_primary_alpha} reports bias and RMSE for the speed
loading parameters $\hat{\alpha}_j$. Bias is small and is falling within $[-0.018, 0.032]$, with no systematic direction. The slight tendency toward positive bias in the $c = 5$ and $c = 10$ conditions is negligible in magnitude and shows no consistent pattern across items. RMSE ranges from approximately 0.031 to 0.093, comparable to that observed for $\hat{\beta}_j$. Neither the position of the boundary parameter $c$ nor the prevalence $\pi$ produces any clear effect
on the speed loading recovery.

\begin{table}[t]
\centering
\caption{Simulation results (primary grid): speed loading $\hat{\alpha}_j$. $N = 256$, $J = 20$.}
\label{tab:sim_primary_alpha}
\small
\resizebox{\textwidth}{!}{%
\begin{tabular}{@{}rrrrrrrrrrrrrrrrrrr@{}}
\toprule
Item & \multicolumn{2}{c}{$c=5$, $\pi=0.15$} & \multicolumn{2}{c}{$c=5$, $\pi=0.25$} & \multicolumn{2}{c}{$c=5$, $\pi=0.4$} & \multicolumn{2}{c}{$c=10$, $\pi=0.15$} & \multicolumn{2}{c}{$c=10$, $\pi=0.25$} & \multicolumn{2}{c}{$c=10$, $\pi=0.4$} & \multicolumn{2}{c}{$c=15$, $\pi=0.15$} & \multicolumn{2}{c}{$c=15$, $\pi=0.25$} & \multicolumn{2}{c}{$c=15$, $\pi=0.4$} \\
\cmidrule(lr){2-3}\cmidrule(lr){4-5}\cmidrule(lr){6-7}\cmidrule(lr){8-9}\cmidrule(lr){10-11}\cmidrule(lr){12-13}\cmidrule(lr){14-15}\cmidrule(lr){16-17}\cmidrule(lr){18-19}
 & Bias & RMSE & Bias & RMSE & Bias & RMSE & Bias & RMSE & Bias & RMSE & Bias & RMSE & Bias & RMSE & Bias & RMSE & Bias & RMSE \\
\midrule
1 & $\phantom{-}0.009$ & $0.067$ & $\phantom{-}0.027$ & $0.081$ & $\phantom{-}0.000$ & $0.044$ & $\phantom{-}0.002$ & $0.078$ & $\phantom{-}0.012$ & $0.044$ & $\phantom{-}0.013$ & $0.043$ & $-0.006$ & $0.037$ & $\phantom{-}0.004$ & $0.031$ & $\phantom{-}0.014$ & $0.059$ \\
2 & $\phantom{-}0.011$ & $0.078$ & $\phantom{-}0.027$ & $0.085$ & $\phantom{-}0.007$ & $0.048$ & $\phantom{-}0.001$ & $0.063$ & $\phantom{-}0.022$ & $0.076$ & $\phantom{-}0.017$ & $0.063$ & $-0.011$ & $0.057$ & $\phantom{-}0.015$ & $0.066$ & $\phantom{-}0.008$ & $0.053$ \\
3 & $\phantom{-}0.007$ & $0.043$ & $\phantom{-}0.019$ & $0.062$ & $\phantom{-}0.005$ & $0.049$ & $\phantom{-}0.000$ & $0.059$ & $\phantom{-}0.016$ & $0.059$ & $\phantom{-}0.023$ & $0.077$ & $-0.011$ & $0.055$ & $\phantom{-}0.013$ & $0.064$ & $\phantom{-}0.006$ & $0.037$ \\
4 & $\phantom{-}0.010$ & $0.059$ & $\phantom{-}0.012$ & $0.038$ & $\phantom{-}0.003$ & $0.039$ & $-0.001$ & $0.063$ & $\phantom{-}0.026$ & $0.090$ & $\phantom{-}0.013$ & $0.044$ & $-0.012$ & $0.055$ & $\phantom{-}0.004$ & $0.034$ & $\phantom{-}0.007$ & $0.043$ \\
5 & $\phantom{-}0.013$ & $0.076$ & $\phantom{-}0.025$ & $0.078$ & $\phantom{-}0.005$ & $0.060$ & $-0.000$ & $0.050$ & $\phantom{-}0.018$ & $0.062$ & $\phantom{-}0.014$ & $0.048$ & $-0.012$ & $0.053$ & $\phantom{-}0.013$ & $0.069$ & $\phantom{-}0.011$ & $0.056$ \\
6 & $\phantom{-}0.014$ & $0.070$ & $\phantom{-}0.010$ & $0.040$ & $\phantom{-}0.007$ & $0.055$ & $\phantom{-}0.001$ & $0.052$ & $\phantom{-}0.011$ & $0.038$ & $\phantom{-}0.019$ & $0.067$ & $-0.018$ & $0.070$ & $\phantom{-}0.015$ & $0.070$ & $\phantom{-}0.010$ & $0.040$ \\
7 & $\phantom{-}0.004$ & $0.036$ & $\phantom{-}0.018$ & $0.062$ & $\phantom{-}0.007$ & $0.062$ & $-0.002$ & $0.039$ & $\phantom{-}0.024$ & $0.081$ & $\phantom{-}0.012$ & $0.040$ & $-0.018$ & $0.076$ & $\phantom{-}0.007$ & $0.035$ & $\phantom{-}0.018$ & $0.084$ \\
8 & $\phantom{-}0.004$ & $0.052$ & $\phantom{-}0.020$ & $0.058$ & $\phantom{-}0.010$ & $0.068$ & $\phantom{-}0.001$ & $0.051$ & $\phantom{-}0.018$ & $0.069$ & $\phantom{-}0.020$ & $0.073$ & $-0.015$ & $0.072$ & $\phantom{-}0.011$ & $0.076$ & $\phantom{-}0.012$ & $0.064$ \\
9 & $\phantom{-}0.009$ & $0.053$ & $\phantom{-}0.017$ & $0.057$ & $\phantom{-}0.007$ & $0.063$ & $\phantom{-}0.002$ & $0.054$ & $\phantom{-}0.019$ & $0.064$ & $\phantom{-}0.014$ & $0.047$ & $-0.010$ & $0.049$ & $\phantom{-}0.012$ & $0.052$ & $\phantom{-}0.014$ & $0.055$ \\
10 & $\phantom{-}0.011$ & $0.064$ & $\phantom{-}0.019$ & $0.054$ & $\phantom{-}0.006$ & $0.061$ & $\phantom{-}0.002$ & $0.054$ & $\phantom{-}0.026$ & $0.093$ & $\phantom{-}0.008$ & $0.034$ & $-0.015$ & $0.073$ & $\phantom{-}0.011$ & $0.064$ & $\phantom{-}0.013$ & $0.075$ \\
11 & $\phantom{-}0.008$ & $0.052$ & $\phantom{-}0.020$ & $0.058$ & $\phantom{-}0.007$ & $0.069$ & $\phantom{-}0.001$ & $0.051$ & $\phantom{-}0.020$ & $0.065$ & $\phantom{-}0.023$ & $0.074$ & $-0.008$ & $0.039$ & $\phantom{-}0.006$ & $0.038$ & $\phantom{-}0.009$ & $0.048$ \\
12 & $\phantom{-}0.007$ & $0.049$ & $\phantom{-}0.028$ & $0.085$ & $\phantom{-}0.005$ & $0.051$ & $\phantom{-}0.002$ & $0.081$ & $\phantom{-}0.016$ & $0.058$ & $\phantom{-}0.020$ & $0.077$ & $-0.012$ & $0.069$ & $\phantom{-}0.012$ & $0.073$ & $\phantom{-}0.019$ & $0.092$ \\
13 & $\phantom{-}0.007$ & $0.042$ & $\phantom{-}0.027$ & $0.083$ & $\phantom{-}0.005$ & $0.067$ & $\phantom{-}0.001$ & $0.036$ & $\phantom{-}0.011$ & $0.039$ & $\phantom{-}0.023$ & $0.067$ & $-0.013$ & $0.062$ & $\phantom{-}0.010$ & $0.067$ & $\phantom{-}0.014$ & $0.093$ \\
14 & $\phantom{-}0.003$ & $0.036$ & $\phantom{-}0.007$ & $0.039$ & $\phantom{-}0.005$ & $0.045$ & $\phantom{-}0.004$ & $0.073$ & $\phantom{-}0.016$ & $0.053$ & $\phantom{-}0.015$ & $0.051$ & $-0.004$ & $0.032$ & $\phantom{-}0.008$ & $0.052$ & $\phantom{-}0.008$ & $0.034$ \\
15 & $\phantom{-}0.010$ & $0.060$ & $\phantom{-}0.011$ & $0.037$ & $\phantom{-}0.002$ & $0.044$ & $\phantom{-}0.003$ & $0.082$ & $\phantom{-}0.015$ & $0.052$ & $\phantom{-}0.016$ & $0.059$ & $-0.015$ & $0.053$ & $\phantom{-}0.010$ & $0.049$ & $\phantom{-}0.018$ & $0.085$ \\
16 & $\phantom{-}0.012$ & $0.075$ & $\phantom{-}0.023$ & $0.067$ & $\phantom{-}0.007$ & $0.050$ & $\phantom{-}0.004$ & $0.050$ & $\phantom{-}0.011$ & $0.057$ & $\phantom{-}0.015$ & $0.043$ & $-0.014$ & $0.073$ & $\phantom{-}0.008$ & $0.042$ & $\phantom{-}0.019$ & $0.081$ \\
17 & $\phantom{-}0.006$ & $0.045$ & $\phantom{-}0.013$ & $0.041$ & $\phantom{-}0.002$ & $0.062$ & $-0.002$ & $0.038$ & $\phantom{-}0.015$ & $0.057$ & $\phantom{-}0.021$ & $0.059$ & $-0.006$ & $0.047$ & $\phantom{-}0.009$ & $0.047$ & $\phantom{-}0.009$ & $0.049$ \\
18 & $\phantom{-}0.009$ & $0.069$ & $\phantom{-}0.032$ & $0.091$ & $\phantom{-}0.005$ & $0.067$ & $\phantom{-}0.002$ & $0.079$ & $\phantom{-}0.011$ & $0.045$ & $\phantom{-}0.017$ & $0.065$ & $-0.013$ & $0.070$ & $\phantom{-}0.012$ & $0.056$ & $\phantom{-}0.009$ & $0.050$ \\
19 & $\phantom{-}0.005$ & $0.040$ & $\phantom{-}0.011$ & $0.050$ & $\phantom{-}0.002$ & $0.073$ & $-0.004$ & $0.071$ & $\phantom{-}0.016$ & $0.058$ & $\phantom{-}0.015$ & $0.049$ & $-0.014$ & $0.066$ & $\phantom{-}0.012$ & $0.067$ & $\phantom{-}0.008$ & $0.056$ \\
20 & $\phantom{-}0.010$ & $0.084$ & $\phantom{-}0.013$ & $0.048$ & $\phantom{-}0.000$ & $0.046$ & $-0.002$ & $0.060$ & $\phantom{-}0.022$ & $0.074$ & $\phantom{-}0.017$ & $0.064$ & $-0.005$ & $0.035$ & $\phantom{-}0.009$ & $0.068$ & $\phantom{-}0.014$ & $0.075$ \\
\bottomrule
\end{tabular}}
\begin{tablenotes}\small
\item Rows correspond to items; columns to simulation conditions. Cells show bias and RMSE for $\alpha_j$. 
\end{tablenotes}
\end{table}

Table~\ref{tab:sim_primary_gamma} reports bias and RMSE for the
change-point effect parameters $\hat{\gamma}_j$, which are the most
directly informative parameters for assessing the model's ability to
identify the post-change change. The pattern here is different from that seen for $\hat{\beta}_j$ and $\hat{\alpha}_j$, as recovery depends on how close the item is to the boundary parameter $c$. Items immediately following the earliest admissible change-point show
the largest negative bias and highest RMSE. Under $c = 5$, item 7
exhibits bias ranging from $-0.125$ to $-0.293$ across the three
prevalence levels, with RMSE between 0.360 and 0.462. This reflects
the fact that items near the boundary are eligible to contribute to the change-point likelihood for a large proportion of respondents, making their $\gamma_j$ estimates sensitive to misclassification of individual change-point locations. As items move further from the boundary, both bias and RMSE decrease substantially and consistently. By items 18--20, bias is negligible and RMSE falls below 0.10 in most conditions, a level comparable to the item-level RMSE observed for $\hat{\beta}_j$.

Prevalence affects recovery of the near-boundary items more than
the distant ones. At item 7, the $\pi = 0.25$ condition yields
considerably smaller bias ($-0.125$) than the $\pi = 0.15$ and
$\pi = 0.40$ conditions ($-0.293$ and $-0.290$ respectively),
suggesting that moderate prevalence provides the most favourable
balance between information about the change-point structure and
stability of the boundary estimates. The $c = 10$, $\pi = 0.15$
condition shows a similar increase in bias and RMSE at item 12,
the first eligible item for that boundary, showing that the
near-boundary pattern is not a  characteristic for a particular condition. Overall, $\hat{\gamma}_j$ recovery is good for items that are not in the immediate proximity of the boundary, and improves with
higher prevalence and longer post-change windows.

\begin{table}[t]
\centering
\caption{Simulation results (primary grid): change-point effect $\hat{\gamma}_j$. $N = 256$, $J = 20$.}
\label{tab:sim_primary_gamma}
\small
\resizebox{\textwidth}{!}{%
\begin{tabular}{@{}rrrrrrrrrrrrrrrrrrr@{}}
\toprule
Item & \multicolumn{2}{c}{$c=5$, $\pi=0.15$} & \multicolumn{2}{c}{$c=5$, $\pi=0.25$} & \multicolumn{2}{c}{$c=5$, $\pi=0.4$} & \multicolumn{2}{c}{$c=10$, $\pi=0.15$} & \multicolumn{2}{c}{$c=10$, $\pi=0.25$} & \multicolumn{2}{c}{$c=10$, $\pi=0.4$} & \multicolumn{2}{c}{$c=15$, $\pi=0.15$} & \multicolumn{2}{c}{$c=15$, $\pi=0.25$} & \multicolumn{2}{c}{$c=15$, $\pi=0.4$} \\
\cmidrule(lr){2-3}\cmidrule(lr){4-5}\cmidrule(lr){6-7}\cmidrule(lr){8-9}\cmidrule(lr){10-11}\cmidrule(lr){12-13}\cmidrule(lr){14-15}\cmidrule(lr){16-17}\cmidrule(lr){18-19}
 & Bias & RMSE & Bias & RMSE & Bias & RMSE & Bias & RMSE & Bias & RMSE & Bias & RMSE & Bias & RMSE & Bias & RMSE & Bias & RMSE \\
\midrule
1 & --- & --- & --- & --- & --- & --- & --- & --- & --- & --- & --- & --- & --- & --- & --- & --- & --- & --- \\
2 & --- & --- & --- & --- & --- & --- & --- & --- & --- & --- & --- & --- & --- & --- & --- & --- & --- & --- \\
3 & --- & --- & --- & --- & --- & --- & --- & --- & --- & --- & --- & --- & --- & --- & --- & --- & --- & --- \\
4 & --- & --- & --- & --- & --- & --- & --- & --- & --- & --- & --- & --- & --- & --- & --- & --- & --- & --- \\
5 & --- & --- & --- & --- & --- & --- & --- & --- & --- & --- & --- & --- & --- & --- & --- & --- & --- & --- \\
6 & --- & --- & --- & --- & --- & --- & --- & --- & --- & --- & --- & --- & --- & --- & --- & --- & --- & --- \\
7 & $-0.293$ & $0.400$ & $-0.125$ & $0.360$ & $-0.290$ & $0.462$ & --- & --- & --- & --- & --- & --- & --- & --- & --- & --- & --- & --- \\
8 & $-0.174$ & $0.331$ & $-0.043$ & $0.271$ & $-0.217$ & $0.305$ & --- & --- & --- & --- & --- & --- & --- & --- & --- & --- & --- & --- \\
9 & $-0.207$ & $0.396$ & $-0.120$ & $0.284$ & $-0.099$ & $0.282$ & --- & --- & --- & --- & --- & --- & --- & --- & --- & --- & --- & --- \\
10 & $-0.079$ & $0.282$ & $-0.051$ & $0.223$ & $-0.072$ & $0.252$ & --- & --- & --- & --- & --- & --- & --- & --- & --- & --- & --- & --- \\
11 & $-0.088$ & $0.267$ & $-0.055$ & $0.183$ & $\phantom{-}0.012$ & $0.162$ & --- & --- & --- & --- & --- & --- & --- & --- & --- & --- & --- & --- \\
12 & $-0.004$ & $0.213$ & $-0.020$ & $0.148$ & $-0.011$ & $0.136$ & $-0.213$ & $0.424$ & $-0.087$ & $0.281$ & $-0.087$ & $0.322$ & --- & --- & --- & --- & --- & --- \\
13 & $-0.028$ & $0.167$ & $-0.004$ & $0.129$ & $-0.014$ & $0.112$ & $-0.126$ & $0.315$ & $-0.060$ & $0.179$ & $-0.004$ & $0.136$ & --- & --- & --- & --- & --- & --- \\
14 & $-0.057$ & $0.212$ & $-0.008$ & $0.103$ & $-0.015$ & $0.118$ & $-0.063$ & $0.230$ & $-0.040$ & $0.137$ & $-0.019$ & $0.113$ & --- & --- & --- & --- & --- & --- \\
15 & $-0.028$ & $0.153$ & $-0.009$ & $0.086$ & $-0.006$ & $0.097$ & $-0.037$ & $0.194$ & $-0.031$ & $0.104$ & $-0.012$ & $0.090$ & --- & --- & --- & --- & --- & --- \\
16 & $-0.021$ & $0.121$ & $-0.000$ & $0.084$ & $-0.005$ & $0.074$ & $\phantom{-}0.012$ & $0.159$ & $\phantom{-}0.002$ & $0.071$ & $-0.010$ & $0.072$ & --- & --- & --- & --- & --- & --- \\
17 & $-0.035$ & $0.114$ & $-0.012$ & $0.076$ & $-0.009$ & $0.068$ & $-0.042$ & $0.167$ & $\phantom{-}0.001$ & $0.080$ & $\phantom{-}0.005$ & $0.064$ & $-0.050$ & $0.204$ & $-0.015$ & $0.135$ & $\phantom{-}0.002$ & $0.121$ \\
18 & $-0.027$ & $0.108$ & $\phantom{-}0.004$ & $0.079$ & $-0.003$ & $0.056$ & $-0.018$ & $0.139$ & $-0.008$ & $0.058$ & $\phantom{-}0.007$ & $0.050$ & $-0.017$ & $0.128$ & $-0.006$ & $0.091$ & $-0.003$ & $0.069$ \\
19 & $-0.033$ & $0.114$ & $\phantom{-}0.004$ & $0.065$ & $-0.007$ & $0.064$ & $-0.016$ & $0.136$ & $\phantom{-}0.006$ & $0.059$ & $\phantom{-}0.013$ & $0.059$ & $\phantom{-}0.001$ & $0.084$ & $-0.003$ & $0.070$ & $\phantom{-}0.001$ & $0.046$ \\
20 & $-0.007$ & $0.095$ & $\phantom{-}0.011$ & $0.060$ & $-0.004$ & $0.063$ & $-0.039$ & $0.126$ & $\phantom{-}0.005$ & $0.052$ & $-0.015$ & $0.065$ & $\phantom{-}0.004$ & $0.068$ & $\phantom{-}0.003$ & $0.054$ & $\phantom{-}0.008$ & $0.050$ \\
\bottomrule
\end{tabular}}
\begin{tablenotes}\small
\item Rows correspond to items; columns to simulation conditions. Cells show bias and RMSE for $\gamma_j$. Dashes (---) indicate items outside the admissible change-point window for that condition ($\gamma_j$ is not defined there).
\end{tablenotes}
\end{table}

Table~\ref{tab:sim_primary_sigma} presents the bias and RMSE for the
residual standard deviation parameters $\hat{\sigma}_j$. The results here differ somewhat from the clear recovery seen for $\hat{\beta}_j$ and $\hat{\alpha}_j$. Bias values are more variable, ranging from
$-0.093$ to $0.129$, and the RMSE tracks the absolute bias
closely throughout. This pattern is
consistent across items and conditions and does not diminish with
increasing prevalence, suggesting it reflects a structural feature of
how residual variance is identified in the marginalised likelihood
rather than finite-sample noise. There is no systematic direction to the bias, and the magnitude is modest, with a large majority falling within $\pm 0.10$. 

\begin{table}[t]
\centering
\caption{Simulation results (primary grid): residual standard deviation $\hat{\sigma}_j$. $N = 256$, $J = 20$.}
\label{tab:sim_primary_sigma}
\small
\resizebox{\textwidth}{!}{%
\begin{tabular}{@{}rrrrrrrrrrrrrrrrrrr@{}}
\toprule
Item & \multicolumn{2}{c}{$c=5$, $\pi=0.15$} & \multicolumn{2}{c}{$c=5$, $\pi=0.25$} & \multicolumn{2}{c}{$c=5$, $\pi=0.4$} & \multicolumn{2}{c}{$c=10$, $\pi=0.15$} & \multicolumn{2}{c}{$c=10$, $\pi=0.25$} & \multicolumn{2}{c}{$c=10$, $\pi=0.4$} & \multicolumn{2}{c}{$c=15$, $\pi=0.15$} & \multicolumn{2}{c}{$c=15$, $\pi=0.25$} & \multicolumn{2}{c}{$c=15$, $\pi=0.4$} \\
\cmidrule(lr){2-3}\cmidrule(lr){4-5}\cmidrule(lr){6-7}\cmidrule(lr){8-9}\cmidrule(lr){10-11}\cmidrule(lr){12-13}\cmidrule(lr){14-15}\cmidrule(lr){16-17}\cmidrule(lr){18-19}
 & Bias & RMSE & Bias & RMSE & Bias & RMSE & Bias & RMSE & Bias & RMSE & Bias & RMSE & Bias & RMSE & Bias & RMSE & Bias & RMSE \\
\midrule
1 & $\phantom{-}0.045$ & $0.048$ & $-0.057$ & $0.058$ & $\phantom{-}0.000$ & $0.016$ & $-0.050$ & $0.052$ & $\phantom{-}0.039$ & $0.042$ & $-0.091$ & $0.092$ & $\phantom{-}0.015$ & $0.021$ & $-0.003$ & $0.012$ & $\phantom{-}0.019$ & $0.024$ \\
2 & $-0.023$ & $0.028$ & $-0.026$ & $0.031$ & $-0.003$ & $0.017$ & $\phantom{-}0.025$ & $0.029$ & $\phantom{-}0.014$ & $0.022$ & $-0.040$ & $0.043$ & $\phantom{-}0.067$ & $0.069$ & $\phantom{-}0.032$ & $0.036$ & $\phantom{-}0.040$ & $0.044$ \\
3 & $-0.001$ & $0.016$ & $\phantom{-}0.081$ & $0.083$ & $\phantom{-}0.104$ & $0.105$ & $\phantom{-}0.006$ & $0.017$ & $\phantom{-}0.053$ & $0.055$ & $\phantom{-}0.061$ & $0.062$ & $\phantom{-}0.081$ & $0.082$ & $\phantom{-}0.026$ & $0.031$ & $-0.087$ & $0.089$ \\
4 & $\phantom{-}0.030$ & $0.034$ & $-0.047$ & $0.049$ & $\phantom{-}0.084$ & $0.085$ & $\phantom{-}0.007$ & $0.018$ & $-0.029$ & $0.033$ & $-0.046$ & $0.047$ & $-0.021$ & $0.025$ & $\phantom{-}0.033$ & $0.036$ & $-0.074$ & $0.075$ \\
5 & $-0.002$ & $0.017$ & $\phantom{-}0.022$ & $0.028$ & $\phantom{-}0.058$ & $0.060$ & $\phantom{-}0.080$ & $0.082$ & $-0.093$ & $0.094$ & $-0.091$ & $0.093$ & $\phantom{-}0.070$ & $0.072$ & $\phantom{-}0.066$ & $0.068$ & $\phantom{-}0.010$ & $0.016$ \\
6 & $\phantom{-}0.016$ & $0.023$ & $\phantom{-}0.034$ & $0.037$ & $\phantom{-}0.002$ & $0.016$ & $-0.003$ & $0.016$ & $\phantom{-}0.051$ & $0.052$ & $-0.019$ & $0.025$ & $\phantom{-}0.034$ & $0.037$ & $\phantom{-}0.019$ & $0.024$ & $-0.044$ & $0.046$ \\
7 & $-0.028$ & $0.031$ & $-0.081$ & $0.083$ & $\phantom{-}0.110$ & $0.111$ & $\phantom{-}0.035$ & $0.038$ & $\phantom{-}0.010$ & $0.018$ & $\phantom{-}0.071$ & $0.072$ & $-0.059$ & $0.061$ & $-0.087$ & $0.088$ & $\phantom{-}0.063$ & $0.065$ \\
8 & $-0.027$ & $0.030$ & $\phantom{-}0.024$ & $0.028$ & $-0.047$ & $0.050$ & $-0.058$ & $0.061$ & $\phantom{-}0.006$ & $0.015$ & $-0.008$ & $0.019$ & $-0.071$ & $0.072$ & $-0.031$ & $0.034$ & $\phantom{-}0.035$ & $0.038$ \\
9 & $\phantom{-}0.013$ & $0.020$ & $\phantom{-}0.070$ & $0.072$ & $-0.022$ & $0.028$ & $\phantom{-}0.045$ & $0.048$ & $-0.042$ & $0.045$ & $\phantom{-}0.007$ & $0.016$ & $\phantom{-}0.033$ & $0.036$ & $-0.071$ & $0.072$ & $\phantom{-}0.088$ & $0.089$ \\
10 & $-0.068$ & $0.070$ & $-0.051$ & $0.053$ & $-0.049$ & $0.052$ & $-0.083$ & $0.084$ & $-0.002$ & $0.018$ & $-0.057$ & $0.059$ & $\phantom{-}0.078$ & $0.080$ & $\phantom{-}0.049$ & $0.051$ & $\phantom{-}0.025$ & $0.030$ \\
11 & $-0.036$ & $0.040$ & $\phantom{-}0.045$ & $0.047$ & $-0.048$ & $0.051$ & $-0.072$ & $0.074$ & $\phantom{-}0.101$ & $0.102$ & $-0.006$ & $0.016$ & $-0.083$ & $0.085$ & $\phantom{-}0.010$ & $0.017$ & $-0.064$ & $0.066$ \\
12 & $-0.011$ & $0.020$ & $\phantom{-}0.001$ & $0.016$ & $\phantom{-}0.022$ & $0.026$ & $-0.073$ & $0.075$ & $-0.064$ & $0.066$ & $-0.071$ & $0.073$ & $\phantom{-}0.014$ & $0.020$ & $\phantom{-}0.023$ & $0.027$ & $\phantom{-}0.090$ & $0.092$ \\
13 & $\phantom{-}0.129$ & $0.130$ & $\phantom{-}0.009$ & $0.020$ & $-0.014$ & $0.022$ & $\phantom{-}0.103$ & $0.104$ & $-0.054$ & $0.056$ & $-0.032$ & $0.035$ & $-0.071$ & $0.073$ & $\phantom{-}0.027$ & $0.031$ & $\phantom{-}0.078$ & $0.079$ \\
14 & $\phantom{-}0.072$ & $0.074$ & $-0.072$ & $0.073$ & $-0.036$ & $0.041$ & $\phantom{-}0.012$ & $0.020$ & $-0.051$ & $0.054$ & $\phantom{-}0.023$ & $0.029$ & $-0.015$ & $0.020$ & $-0.016$ & $0.021$ & $-0.037$ & $0.040$ \\
15 & $-0.044$ & $0.048$ & $-0.048$ & $0.051$ & $\phantom{-}0.053$ & $0.056$ & $-0.029$ & $0.034$ & $\phantom{-}0.038$ & $0.041$ & $\phantom{-}0.081$ & $0.083$ & $-0.017$ & $0.022$ & $-0.016$ & $0.021$ & $\phantom{-}0.001$ & $0.015$ \\
16 & $-0.053$ & $0.057$ & $\phantom{-}0.000$ & $0.015$ & $-0.019$ & $0.024$ & $\phantom{-}0.081$ & $0.083$ & $\phantom{-}0.013$ & $0.020$ & $\phantom{-}0.101$ & $0.102$ & $\phantom{-}0.035$ & $0.038$ & $-0.006$ & $0.014$ & $\phantom{-}0.078$ & $0.079$ \\
17 & $\phantom{-}0.084$ & $0.086$ & $\phantom{-}0.005$ & $0.016$ & $-0.073$ & $0.075$ & $-0.039$ & $0.044$ & $\phantom{-}0.052$ & $0.055$ & $\phantom{-}0.021$ & $0.030$ & $\phantom{-}0.064$ & $0.066$ & $\phantom{-}0.078$ & $0.080$ & $-0.067$ & $0.069$ \\
18 & $-0.055$ & $0.058$ & $\phantom{-}0.075$ & $0.077$ & $-0.060$ & $0.063$ & $\phantom{-}0.024$ & $0.031$ & $-0.048$ & $0.050$ & $\phantom{-}0.012$ & $0.024$ & $-0.012$ & $0.019$ & $-0.083$ & $0.084$ & $-0.062$ & $0.065$ \\
19 & $\phantom{-}0.020$ & $0.028$ & $\phantom{-}0.046$ & $0.049$ & $\phantom{-}0.062$ & $0.066$ & $\phantom{-}0.103$ & $0.105$ & $\phantom{-}0.010$ & $0.019$ & $\phantom{-}0.065$ & $0.067$ & $-0.060$ & $0.062$ & $\phantom{-}0.017$ & $0.024$ & $-0.080$ & $0.083$ \\
20 & $-0.041$ & $0.048$ & $-0.005$ & $0.021$ & $-0.075$ & $0.077$ & $-0.063$ & $0.067$ & $\phantom{-}0.039$ & $0.044$ & $\phantom{-}0.079$ & $0.082$ & $-0.035$ & $0.040$ & $-0.041$ & $0.047$ & $\phantom{-}0.058$ & $0.064$ \\
\bottomrule
\end{tabular}}
\begin{tablenotes}\small
\item Rows correspond to items; columns to simulation conditions. Cells show bias and RMSE for $\sigma_j$. 
\end{tablenotes}
\end{table}

\subsubsection{Secondary grid}

Table~\ref{tab:sim_secondary_cp} presents the change-point recovery and structural parameter estimates for the secondary grid. The clearest
finding is that test length improves $\tau$ recovery more reliably than
sample size. The best overall performance is achieved at $J = 40$,
$N = 600$, with MAE(mode) $= 0.176$, while increasing $N$ from 200
to 1800 at fixed $J = 20$ reduces MAE(mode) only from 0.278 to 0.241.
Since a longer test provides more post-change items per respondent, these results have a rather natural explanation.

The relationship between $N$ and $\tau$ recovery is non-monotone. At $J = 30$ and $J = 40$, the $N = 1800$ condition performs no better
than $N = 600$ and in some cases slightly worse. The corresponding structural parameter RMSEs do however improve with $N$ as expected, most clearly for $\psi_3$ (0.396, 0.161, 0.098 for $J = 30$). The non-monotonicity in MAE($\hat{\tau}$) likely reflects that tau recovery at $\pi = 0.15$ is ultimately limited by the information available per changer rather than by aggregate sample size, and once $N$ is large enough to estimate the structural parameters well, further increases yield smaller improvements for individual-level classification.

Structural parameter bias is small throughout. The main exception is the RMSE for
$\psi_2$, which is elevated at $N = 200$ (0.517 for $J = 20$,
0.525 for $J = 30$) and decreases clearly with $N$, consistent with
the no-change probability being difficult to estimate precisely when
changers are rare and the sample is small. The $\psi_3$ RMSE follows
the same pattern, halving or better between $N = 200$ and $N = 1800$
across all test lengths.

\begin{table}[t]
\centering
\caption{Simulation results (secondary grid): change-point recovery and structural parameters. $\pi = 0.15$ fixed, $c \in \{12, 18, 24\}$ matched to $J$; rows vary $N$ and $J$.}
\label{tab:sim_secondary_cp}
\small
\resizebox{\textwidth}{!}{%
\begin{tabular}{@{}lrr rr rr rr@{}}
\toprule
Condition & \multicolumn{2}{c}{MAE($\hat{\tau}$)} & \multicolumn{2}{c}{$\psi_1$} & \multicolumn{2}{c}{$\psi_2$} & \multicolumn{2}{c}{$\psi_3$} \\
\cmidrule(lr){2-3}\cmidrule(lr){4-5} \cmidrule(lr){6-7}\cmidrule(lr){8-9}
 & Mode & Mean & Bias & RMSE & Bias & RMSE & Bias & RMSE \\
\midrule
$N=200$, $J=20$ & $0.278$ & $0.342$ & $-0.079$ & $0.264$ & $-0.022$ & $0.517$ & $-0.041$ & $0.361$ \\
$N=600$, $J=20$ & $0.244$ & $0.313$ & $-0.021$ & $0.099$ & $-0.065$ & $0.233$ & $\phantom{-}0.015$ & $0.193$ \\
$N=1800$, $J=20$ & $0.241$ & $0.339$ & $\phantom{-}0.010$ & $0.077$ & $-0.109$ & $0.249$ & $-0.005$ & $0.132$ \\
$N=200$, $J=30$ & $0.352$ & $0.407$ & $-0.037$ & $0.179$ & $-0.102$ & $0.525$ & $\phantom{-}0.076$ & $0.396$ \\
$N=600$, $J=30$ & $0.220$ & $0.266$ & $-0.026$ & $0.140$ & $-0.012$ & $0.173$ & $-0.011$ & $0.161$ \\
$N=1800$, $J=30$ & $0.242$ & $0.299$ & $-0.020$ & $0.193$ & $-0.095$ & $0.178$ & $\phantom{-}0.053$ & $0.098$ \\
$N=200$, $J=40$ & $0.262$ & $0.310$ & $-0.027$ & $0.108$ & $\phantom{-}0.020$ & $0.325$ & $-0.051$ & $0.291$ \\ 
$N=600$, $J=40$ & $0.176$ & $0.216$ & $-0.017$ & $0.153$ & $\phantom{-}0.001$ & $0.185$ & $-0.021$ & $0.137$ \\
$N=1800$, $J=40$ & $0.263$ & $0.320$ & $-0.022$ & $0.142$ & $-0.066$ & $0.143$ & $\phantom{-}0.044$ & $0.084$ \\
\bottomrule
\end{tabular}}
\begin{tablenotes}\small
\item MAE($\hat{\tau}$): mean absolute error of the modal (Mode) and posterior mean (Mean) change-point estimate. True values: $\psi_1 = 0.2$, $\psi_3 = -0.5$; $\psi_2$ set per condition to match target prevalence $\pi$. 
\end{tablenotes}
\end{table}

Table~\ref{tab:sim_secondary_beta} reports bias and RMSE for
$\hat{\beta}_j$ across all items and conditions in the secondary grid.
The results align with the conclusion from the primary grid: recovery is
uniformly good regardless of test length or sample size. Bias is negligible throughout, with values almost entirely within $[-0.016, 0.036]$. 

\begin{table}[t]
\centering
\caption{Simulation results (secondary grid): item time intensity $\hat{\beta}_j$. $\pi = 0.15$, $c$ matched to $J$.}
\label{tab:sim_secondary_beta}
\small
\resizebox{\textwidth}{!}{%
\begin{tabular}{@{}rrrrrrrrrrrrrrrrrrr@{}}
\toprule
Item & \multicolumn{2}{c}{$N=200$, $J=20$} & \multicolumn{2}{c}{$N=600$, $J=20$} & \multicolumn{2}{c}{$N=1800$, $J=20$} & \multicolumn{2}{c}{$N=200$, $J=30$} & \multicolumn{2}{c}{$N=600$, $J=30$} & \multicolumn{2}{c}{$N=1800$, $J=30$} & \multicolumn{2}{c}{$N=200$, $J=40$} & \multicolumn{2}{c}{$N=600$, $J=40$} & \multicolumn{2}{c}{$N=1800$, $J=40$} \\
\cmidrule(lr){2-3}\cmidrule(lr){4-5}\cmidrule(lr){6-7}\cmidrule(lr){8-9}\cmidrule(lr){10-11}\cmidrule(lr){12-13}\cmidrule(lr){14-15}\cmidrule(lr){16-17}\cmidrule(lr){18-19}
 & Bias & RMSE & Bias & RMSE & Bias & RMSE & Bias & RMSE & Bias & RMSE & Bias & RMSE & Bias & RMSE & Bias & RMSE & Bias & RMSE \\
\midrule
1 & $\phantom{-}0.027$ & $0.137$ & $\phantom{-}0.002$ & $0.049$ & $-0.007$ & $0.024$ & $-0.000$ & $0.063$ & $-0.004$ & $0.095$ & $\phantom{-}0.001$ & $0.062$ & $\phantom{-}0.019$ & $0.066$ & $\phantom{-}0.010$ & $0.066$ & $\phantom{-}0.022$ & $0.082$ \\
2 & $\phantom{-}0.022$ & $0.107$ & $\phantom{-}0.003$ & $0.045$ & $-0.013$ & $0.043$ & $\phantom{-}0.001$ & $0.117$ & $-0.007$ & $0.106$ & $\phantom{-}0.002$ & $0.048$ & $\phantom{-}0.021$ & $0.100$ & $\phantom{-}0.015$ & $0.089$ & $\phantom{-}0.013$ & $0.051$ \\
3 & $\phantom{-}0.024$ & $0.125$ & $\phantom{-}0.003$ & $0.045$ & $-0.005$ & $0.023$ & $-0.001$ & $0.105$ & $-0.003$ & $0.069$ & $\phantom{-}0.000$ & $0.077$ & $\phantom{-}0.025$ & $0.112$ & $\phantom{-}0.015$ & $0.095$ & $\phantom{-}0.015$ & $0.061$ \\
4 & $\phantom{-}0.013$ & $0.061$ & $\phantom{-}0.003$ & $0.049$ & $-0.006$ & $0.021$ & $-0.002$ & $0.076$ & $-0.002$ & $0.043$ & $-0.001$ & $0.072$ & $\phantom{-}0.027$ & $0.106$ & $\phantom{-}0.010$ & $0.061$ & $\phantom{-}0.014$ & $0.052$ \\
5 & $\phantom{-}0.030$ & $0.143$ & $\phantom{-}0.006$ & $0.080$ & $-0.009$ & $0.032$ & $-0.003$ & $0.101$ & $-0.006$ & $0.068$ & $\phantom{-}0.000$ & $0.038$ & $\phantom{-}0.016$ & $0.074$ & $\phantom{-}0.010$ & $0.067$ & $\phantom{-}0.019$ & $0.075$ \\
6 & $\phantom{-}0.016$ & $0.075$ & $\phantom{-}0.003$ & $0.046$ & $-0.006$ & $0.022$ & $-0.004$ & $0.089$ & $-0.004$ & $0.107$ & $\phantom{-}0.001$ & $0.048$ & $\phantom{-}0.022$ & $0.101$ & $\phantom{-}0.006$ & $0.036$ & $\phantom{-}0.014$ & $0.052$ \\
7 & $\phantom{-}0.014$ & $0.072$ & $\phantom{-}0.003$ & $0.064$ & $-0.015$ & $0.050$ & $-0.003$ & $0.083$ & $-0.003$ & $0.058$ & $-0.001$ & $0.063$ & $\phantom{-}0.028$ & $0.121$ & $\phantom{-}0.014$ & $0.059$ & $\phantom{-}0.019$ & $0.072$ \\
8 & $\phantom{-}0.024$ & $0.116$ & $\phantom{-}0.003$ & $0.044$ & $-0.010$ & $0.035$ & $\phantom{-}0.003$ & $0.071$ & $-0.005$ & $0.073$ & $\phantom{-}0.001$ & $0.041$ & $\phantom{-}0.014$ & $0.070$ & $\phantom{-}0.010$ & $0.055$ & $\phantom{-}0.020$ & $0.079$ \\
9 & $\phantom{-}0.028$ & $0.140$ & $\phantom{-}0.006$ & $0.069$ & $-0.007$ & $0.028$ & $-0.004$ & $0.114$ & $-0.004$ & $0.058$ & $-0.000$ & $0.072$ & $\phantom{-}0.017$ & $0.073$ & $\phantom{-}0.017$ & $0.089$ & $\phantom{-}0.017$ & $0.068$ \\
10 & $\phantom{-}0.024$ & $0.134$ & $\phantom{-}0.004$ & $0.062$ & $-0.013$ & $0.048$ & $\phantom{-}0.000$ & $0.122$ & $-0.002$ & $0.042$ & $-0.001$ & $0.057$ & $\phantom{-}0.036$ & $0.139$ & $\phantom{-}0.005$ & $0.034$ & $\phantom{-}0.014$ & $0.053$ \\
11 & $\phantom{-}0.022$ & $0.107$ & $\phantom{-}0.001$ & $0.033$ & $-0.015$ & $0.050$ & $-0.003$ & $0.092$ & $-0.004$ & $0.081$ & $-0.001$ & $0.036$ & $\phantom{-}0.026$ & $0.133$ & $\phantom{-}0.011$ & $0.067$ & $\phantom{-}0.014$ & $0.056$ \\
12 & $\phantom{-}0.013$ & $0.066$ & $\phantom{-}0.003$ & $0.056$ & $-0.011$ & $0.044$ & $-0.004$ & $0.048$ & $-0.004$ & $0.065$ & $\phantom{-}0.001$ & $0.033$ & $\phantom{-}0.025$ & $0.112$ & $\phantom{-}0.013$ & $0.079$ & $\phantom{-}0.018$ & $0.067$ \\
13 & $\phantom{-}0.008$ & $0.059$ & $\phantom{-}0.006$ & $0.053$ & $-0.013$ & $0.048$ & $-0.002$ & $0.119$ & $-0.001$ & $0.049$ & $\phantom{-}0.001$ & $0.041$ & $\phantom{-}0.019$ & $0.089$ & $\phantom{-}0.017$ & $0.094$ & $\phantom{-}0.013$ & $0.052$ \\
14 & $\phantom{-}0.013$ & $0.075$ & $\phantom{-}0.004$ & $0.053$ & $-0.008$ & $0.029$ & $-0.005$ & $0.091$ & $-0.002$ & $0.042$ & $\phantom{-}0.001$ & $0.065$ & $\phantom{-}0.020$ & $0.104$ & $\phantom{-}0.008$ & $0.051$ & $\phantom{-}0.012$ & $0.047$ \\
15 & $\phantom{-}0.021$ & $0.120$ & $\phantom{-}0.001$ & $0.048$ & $-0.010$ & $0.032$ & $-0.003$ & $0.112$ & $-0.005$ & $0.073$ & $\phantom{-}0.000$ & $0.063$ & $\phantom{-}0.025$ & $0.106$ & $\phantom{-}0.015$ & $0.085$ & $\phantom{-}0.016$ & $0.061$ \\
16 & $\phantom{-}0.015$ & $0.094$ & $\phantom{-}0.002$ & $0.066$ & $-0.008$ & $0.022$ & $-0.003$ & $0.108$ & $-0.004$ & $0.094$ & $-0.002$ & $0.066$ & $\phantom{-}0.015$ & $0.068$ & $\phantom{-}0.014$ & $0.078$ & $\phantom{-}0.010$ & $0.039$ \\
17 & $\phantom{-}0.024$ & $0.137$ & $-0.001$ & $0.035$ & $-0.010$ & $0.030$ & $-0.001$ & $0.074$ & $-0.005$ & $0.077$ & $\phantom{-}0.000$ & $0.068$ & $\phantom{-}0.023$ & $0.100$ & $\phantom{-}0.010$ & $0.051$ & $\phantom{-}0.015$ & $0.061$ \\
18 & $\phantom{-}0.022$ & $0.119$ & $\phantom{-}0.002$ & $0.067$ & $-0.014$ & $0.043$ & $\phantom{-}0.000$ & $0.051$ & $-0.005$ & $0.070$ & $\phantom{-}0.000$ & $0.045$ & $\phantom{-}0.020$ & $0.086$ & $\phantom{-}0.012$ & $0.078$ & $\phantom{-}0.020$ & $0.073$ \\
19 & $\phantom{-}0.005$ & $0.060$ & $-0.000$ & $0.064$ & $-0.016$ & $0.042$ & $-0.003$ & $0.108$ & $-0.006$ & $0.109$ & $-0.000$ & $0.051$ & $\phantom{-}0.032$ & $0.126$ & $\phantom{-}0.010$ & $0.056$ & $\phantom{-}0.008$ & $0.032$ \\
20 & $\phantom{-}0.021$ & $0.141$ & $-0.001$ & $0.071$ & $-0.010$ & $0.024$ & $-0.002$ & $0.071$ & $-0.003$ & $0.048$ & $-0.000$ & $0.037$ & $\phantom{-}0.030$ & $0.130$ & $\phantom{-}0.013$ & $0.077$ & $\phantom{-}0.012$ & $0.045$ \\
21 & --- & --- & --- & --- & --- & --- & $-0.008$ & $0.121$ & $-0.005$ & $0.079$ & $-0.002$ & $0.071$ & $\phantom{-}0.025$ & $0.102$ & $\phantom{-}0.010$ & $0.050$ & $\phantom{-}0.010$ & $0.043$ \\
22 & --- & --- & --- & --- & --- & --- & $-0.003$ & $0.087$ & $-0.004$ & $0.051$ & $-0.002$ & $0.035$ & $\phantom{-}0.016$ & $0.069$ & $\phantom{-}0.006$ & $0.048$ & $\phantom{-}0.019$ & $0.077$ \\
23 & --- & --- & --- & --- & --- & --- & $-0.000$ & $0.106$ & $-0.005$ & $0.098$ & $-0.001$ & $0.055$ & $\phantom{-}0.018$ & $0.091$ & $\phantom{-}0.010$ & $0.067$ & $\phantom{-}0.008$ & $0.030$ \\
24 & --- & --- & --- & --- & --- & --- & $-0.007$ & $0.076$ & $-0.003$ & $0.043$ & $-0.000$ & $0.050$ & $\phantom{-}0.033$ & $0.128$ & $\phantom{-}0.012$ & $0.065$ & $\phantom{-}0.013$ & $0.050$ \\
25 & --- & --- & --- & --- & --- & --- & $-0.001$ & $0.053$ & $-0.004$ & $0.039$ & $-0.001$ & $0.059$ & $\phantom{-}0.019$ & $0.076$ & $\phantom{-}0.012$ & $0.077$ & $\phantom{-}0.008$ & $0.030$ \\
26 & --- & --- & --- & --- & --- & --- & $-0.007$ & $0.088$ & $-0.004$ & $0.045$ & $-0.003$ & $0.036$ & $\phantom{-}0.011$ & $0.054$ & $\phantom{-}0.013$ & $0.082$ & $\phantom{-}0.015$ & $0.058$ \\
27 & --- & --- & --- & --- & --- & --- & $-0.006$ & $0.114$ & $-0.002$ & $0.044$ & $-0.002$ & $0.074$ & $\phantom{-}0.013$ & $0.054$ & $\phantom{-}0.005$ & $0.041$ & $\phantom{-}0.021$ & $0.083$ \\
28 & --- & --- & --- & --- & --- & --- & $-0.010$ & $0.073$ & $-0.007$ & $0.098$ & $-0.004$ & $0.054$ & $\phantom{-}0.030$ & $0.130$ & $\phantom{-}0.008$ & $0.054$ & $\phantom{-}0.020$ & $0.083$ \\
29 & --- & --- & --- & --- & --- & --- & $-0.007$ & $0.080$ & $-0.004$ & $0.073$ & $-0.005$ & $0.032$ & $\phantom{-}0.026$ & $0.107$ & $\phantom{-}0.015$ & $0.081$ & $\phantom{-}0.020$ & $0.077$ \\
30 & --- & --- & --- & --- & --- & --- & $-0.012$ & $0.098$ & $-0.005$ & $0.049$ & $-0.005$ & $0.066$ & $\phantom{-}0.029$ & $0.126$ & $\phantom{-}0.015$ & $0.084$ & $\phantom{-}0.018$ & $0.070$ \\
31 & --- & --- & --- & --- & --- & --- & --- & --- & --- & --- & --- & --- & $\phantom{-}0.023$ & $0.110$ & $\phantom{-}0.006$ & $0.042$ & $\phantom{-}0.012$ & $0.048$ \\
32 & --- & --- & --- & --- & --- & --- & --- & --- & --- & --- & --- & --- & $\phantom{-}0.013$ & $0.076$ & $\phantom{-}0.014$ & $0.084$ & $\phantom{-}0.008$ & $0.037$ \\
33 & --- & --- & --- & --- & --- & --- & --- & --- & --- & --- & --- & --- & $\phantom{-}0.008$ & $0.057$ & $\phantom{-}0.015$ & $0.088$ & $\phantom{-}0.010$ & $0.039$ \\
34 & --- & --- & --- & --- & --- & --- & --- & --- & --- & --- & --- & --- & $\phantom{-}0.023$ & $0.110$ & $\phantom{-}0.006$ & $0.049$ & $\phantom{-}0.015$ & $0.060$ \\
35 & --- & --- & --- & --- & --- & --- & --- & --- & --- & --- & --- & --- & $\phantom{-}0.013$ & $0.069$ & $\phantom{-}0.013$ & $0.078$ & $\phantom{-}0.017$ & $0.072$ \\
36 & --- & --- & --- & --- & --- & --- & --- & --- & --- & --- & --- & --- & $\phantom{-}0.023$ & $0.107$ & $\phantom{-}0.004$ & $0.044$ & $\phantom{-}0.021$ & $0.084$ \\
37 & --- & --- & --- & --- & --- & --- & --- & --- & --- & --- & --- & --- & $\phantom{-}0.011$ & $0.070$ & $\phantom{-}0.009$ & $0.065$ & $\phantom{-}0.009$ & $0.044$ \\
38 & --- & --- & --- & --- & --- & --- & --- & --- & --- & --- & --- & --- & $\phantom{-}0.017$ & $0.099$ & $\phantom{-}0.008$ & $0.058$ & $\phantom{-}0.018$ & $0.081$ \\
39 & --- & --- & --- & --- & --- & --- & --- & --- & --- & --- & --- & --- & $\phantom{-}0.023$ & $0.109$ & $\phantom{-}0.011$ & $0.079$ & $\phantom{-}0.005$ & $0.034$ \\
40 & --- & --- & --- & --- & --- & --- & --- & --- & --- & --- & --- & --- & $\phantom{-}0.018$ & $0.095$ & $\phantom{-}0.003$ & $0.046$ & $\phantom{-}0.005$ & $0.028$ \\
\bottomrule
\end{tabular}}
\begin{tablenotes}\small
\item Rows correspond to items; columns to simulation conditions. Cells show bias and RMSE for $\beta_j$. Dashes (---) indicate items outside the admissible change-point window for that condition ($\gamma_j$ is not defined there).
\end{tablenotes}
\end{table}

Table~\ref{tab:sim_secondary_alpha} presents the bias and RMSE for the loading parameters $\hat{\alpha}_j$ in the secondary grid. The recovery is good, with bias magnitudes rarely exceeding 0.05 in absolute value, and the RMSE values shrink with $N$ as expected. The RMSE values are around 0.04--0.09 at $N = 200$ and fall to 0.015--0.035 at $N = 1800$. Unlike $\hat{\beta}_j$, there is
no meaningful dependence on item position or test length, and the
extension to $J = 30$ and $J = 40$ introduces no additional complexity
in the recovery of loadings for the new items.
\begin{table}[t]
\centering
\caption{Simulation results (secondary grid): speed loading $\hat{\alpha}_j$. $\pi = 0.15$, $c$ matched to $J$.}
\label{tab:sim_secondary_alpha}
\small
\resizebox{\textwidth}{!}{%
\begin{tabular}{@{}rrrrrrrrrrrrrrrrrrr@{}}
\toprule
Item & \multicolumn{2}{c}{$N=200$, $J=20$} & \multicolumn{2}{c}{$N=600$, $J=20$} & \multicolumn{2}{c}{$N=1800$, $J=20$} & \multicolumn{2}{c}{$N=200$, $J=30$} & \multicolumn{2}{c}{$N=600$, $J=30$} & \multicolumn{2}{c}{$N=1800$, $J=30$} & \multicolumn{2}{c}{$N=200$, $J=40$} & \multicolumn{2}{c}{$N=600$, $J=40$} & \multicolumn{2}{c}{$N=1800$, $J=40$} \\
\cmidrule(lr){2-3}\cmidrule(lr){4-5}\cmidrule(lr){6-7}\cmidrule(lr){8-9}\cmidrule(lr){10-11}\cmidrule(lr){12-13}\cmidrule(lr){14-15}\cmidrule(lr){16-17}\cmidrule(lr){18-19}
 & Bias & RMSE & Bias & RMSE & Bias & RMSE & Bias & RMSE & Bias & RMSE & Bias & RMSE & Bias & RMSE & Bias & RMSE & Bias & RMSE \\
\midrule
1 & $-0.021$ & $0.082$ & $\phantom{-}0.007$ & $0.041$ & $\phantom{-}0.003$ & $0.017$ & $-0.006$ & $0.042$ & $\phantom{-}0.042$ & $0.089$ & $-0.015$ & $0.050$ & $\phantom{-}0.008$ & $0.038$ & $\phantom{-}0.006$ & $0.051$ & $-0.050$ & $0.080$ \\
2 & $-0.015$ & $0.064$ & $\phantom{-}0.005$ & $0.035$ & $\phantom{-}0.005$ & $0.029$ & $-0.004$ & $0.072$ & $\phantom{-}0.045$ & $0.101$ & $-0.013$ & $0.039$ & $\phantom{-}0.016$ & $0.062$ & $\phantom{-}0.006$ & $0.075$ & $-0.031$ & $0.050$ \\
3 & $-0.018$ & $0.071$ & $\phantom{-}0.005$ & $0.037$ & $\phantom{-}0.003$ & $0.017$ & $-0.002$ & $0.063$ & $\phantom{-}0.029$ & $0.064$ & $-0.019$ & $0.062$ & $\phantom{-}0.012$ & $0.068$ & $\phantom{-}0.008$ & $0.077$ & $-0.037$ & $0.060$ \\
4 & $-0.010$ & $0.041$ & $\phantom{-}0.004$ & $0.038$ & $\phantom{-}0.002$ & $0.015$ & $\phantom{-}0.002$ & $0.055$ & $\phantom{-}0.015$ & $0.038$ & $-0.018$ & $0.058$ & $\phantom{-}0.017$ & $0.066$ & $\phantom{-}0.001$ & $0.048$ & $-0.030$ & $0.049$ \\
5 & $-0.021$ & $0.084$ & $\phantom{-}0.009$ & $0.063$ & $\phantom{-}0.005$ & $0.022$ & $\phantom{-}0.001$ & $0.065$ & $\phantom{-}0.027$ & $0.063$ & $-0.009$ & $0.031$ & $\phantom{-}0.008$ & $0.047$ & $\phantom{-}0.007$ & $0.054$ & $-0.047$ & $0.074$ \\
6 & $-0.011$ & $0.046$ & $\phantom{-}0.005$ & $0.037$ & $\phantom{-}0.002$ & $0.015$ & $-0.003$ & $0.055$ & $\phantom{-}0.047$ & $0.100$ & $-0.012$ & $0.038$ & $\phantom{-}0.014$ & $0.065$ & $\phantom{-}0.001$ & $0.028$ & $-0.032$ & $0.051$ \\
7 & $-0.009$ & $0.044$ & $\phantom{-}0.007$ & $0.049$ & $\phantom{-}0.006$ & $0.033$ & $-0.001$ & $0.049$ & $\phantom{-}0.024$ & $0.054$ & $-0.016$ & $0.050$ & $\phantom{-}0.017$ & $0.075$ & $\phantom{-}0.001$ & $0.048$ & $-0.045$ & $0.072$ \\
8 & $-0.015$ & $0.068$ & $\phantom{-}0.005$ & $0.035$ & $\phantom{-}0.005$ & $0.024$ & $-0.001$ & $0.047$ & $\phantom{-}0.030$ & $0.069$ & $-0.012$ & $0.035$ & $\phantom{-}0.017$ & $0.050$ & $\phantom{-}0.005$ & $0.041$ & $-0.048$ & $0.076$ \\
9 & $-0.023$ & $0.086$ & $\phantom{-}0.008$ & $0.054$ & $\phantom{-}0.004$ & $0.021$ & $-0.003$ & $0.069$ & $\phantom{-}0.025$ & $0.055$ & $-0.019$ & $0.059$ & $\phantom{-}0.008$ & $0.045$ & $\phantom{-}0.004$ & $0.070$ & $-0.042$ & $0.067$ \\
10 & $-0.023$ & $0.079$ & $\phantom{-}0.007$ & $0.047$ & $\phantom{-}0.007$ & $0.032$ & $-0.003$ & $0.075$ & $\phantom{-}0.017$ & $0.042$ & $-0.015$ & $0.046$ & $\phantom{-}0.024$ & $0.086$ & $\phantom{-}0.004$ & $0.030$ & $-0.033$ & $0.053$ \\
11 & $-0.016$ & $0.066$ & $\phantom{-}0.003$ & $0.024$ & $\phantom{-}0.006$ & $0.035$ & $\phantom{-}0.001$ & $0.059$ & $\phantom{-}0.033$ & $0.072$ & $-0.010$ & $0.029$ & $\phantom{-}0.016$ & $0.080$ & $\phantom{-}0.005$ & $0.056$ & $-0.034$ & $0.055$ \\
12 & $-0.012$ & $0.043$ & $\phantom{-}0.006$ & $0.043$ & $\phantom{-}0.005$ & $0.030$ & $-0.002$ & $0.036$ & $\phantom{-}0.028$ & $0.061$ & $-0.008$ & $0.025$ & $\phantom{-}0.013$ & $0.067$ & $\phantom{-}0.006$ & $0.064$ & $-0.041$ & $0.065$ \\
13 & $-0.008$ & $0.037$ & $\phantom{-}0.004$ & $0.041$ & $\phantom{-}0.007$ & $0.033$ & $-0.003$ & $0.071$ & $\phantom{-}0.020$ & $0.047$ & $-0.011$ & $0.034$ & $\phantom{-}0.011$ & $0.053$ & $\phantom{-}0.006$ & $0.078$ & $-0.031$ & $0.049$ \\
14 & $-0.014$ & $0.050$ & $\phantom{-}0.007$ & $0.042$ & $\phantom{-}0.003$ & $0.021$ & $-0.004$ & $0.058$ & $\phantom{-}0.020$ & $0.041$ & $-0.018$ & $0.054$ & $\phantom{-}0.011$ & $0.059$ & $\phantom{-}0.003$ & $0.041$ & $-0.027$ & $0.045$ \\
15 & $-0.018$ & $0.070$ & $\phantom{-}0.005$ & $0.038$ & $\phantom{-}0.004$ & $0.025$ & $-0.007$ & $0.073$ & $\phantom{-}0.032$ & $0.068$ & $-0.017$ & $0.052$ & $\phantom{-}0.013$ & $0.064$ & $\phantom{-}0.007$ & $0.070$ & $-0.037$ & $0.059$ \\
16 & $-0.015$ & $0.059$ & $\phantom{-}0.008$ & $0.051$ & $\phantom{-}0.004$ & $0.015$ & $-0.000$ & $0.065$ & $\phantom{-}0.041$ & $0.087$ & $-0.018$ & $0.053$ & $\phantom{-}0.007$ & $0.040$ & $\phantom{-}0.004$ & $0.064$ & $-0.022$ & $0.037$ \\
17 & $-0.024$ & $0.080$ & $\phantom{-}0.005$ & $0.028$ & $\phantom{-}0.004$ & $0.023$ & $\phantom{-}0.003$ & $0.049$ & $\phantom{-}0.032$ & $0.069$ & $-0.017$ & $0.055$ & $\phantom{-}0.018$ & $0.064$ & $\phantom{-}0.004$ & $0.041$ & $-0.037$ & $0.059$ \\
18 & $-0.016$ & $0.074$ & $\phantom{-}0.008$ & $0.056$ & $\phantom{-}0.006$ & $0.029$ & $\phantom{-}0.003$ & $0.034$ & $\phantom{-}0.031$ & $0.067$ & $-0.012$ & $0.038$ & $\phantom{-}0.012$ & $0.056$ & $\phantom{-}0.004$ & $0.062$ & $-0.043$ & $0.069$ \\
19 & $-0.007$ & $0.037$ & $\phantom{-}0.009$ & $0.050$ & $\phantom{-}0.004$ & $0.028$ & $-0.002$ & $0.067$ & $\phantom{-}0.045$ & $0.100$ & $-0.013$ & $0.041$ & $\phantom{-}0.014$ & $0.076$ & $\phantom{-}0.002$ & $0.043$ & $-0.019$ & $0.030$ \\
20 & $-0.020$ & $0.081$ & $\phantom{-}0.009$ & $0.058$ & $\phantom{-}0.003$ & $0.017$ & $-0.003$ & $0.049$ & $\phantom{-}0.018$ & $0.043$ & $-0.010$ & $0.029$ & $\phantom{-}0.014$ & $0.079$ & $\phantom{-}0.007$ & $0.062$ & $-0.027$ & $0.043$ \\
21 & --- & --- & --- & --- & --- & --- & $-0.002$ & $0.076$ & $\phantom{-}0.035$ & $0.076$ & $-0.019$ & $0.058$ & $\phantom{-}0.014$ & $0.067$ & $\phantom{-}0.002$ & $0.043$ & $-0.024$ & $0.040$ \\
22 & --- & --- & --- & --- & --- & --- & $-0.001$ & $0.051$ & $\phantom{-}0.019$ & $0.047$ & $-0.009$ & $0.029$ & $\phantom{-}0.012$ & $0.046$ & $\phantom{-}0.004$ & $0.041$ & $-0.045$ & $0.074$ \\
23 & --- & --- & --- & --- & --- & --- & $-0.006$ & $0.068$ & $\phantom{-}0.041$ & $0.091$ & $-0.014$ & $0.045$ & $\phantom{-}0.011$ & $0.061$ & $\phantom{-}0.005$ & $0.053$ & $-0.018$ & $0.029$ \\
24 & --- & --- & --- & --- & --- & --- & $-0.003$ & $0.047$ & $\phantom{-}0.021$ & $0.042$ & $-0.012$ & $0.041$ & $\phantom{-}0.020$ & $0.076$ & $\phantom{-}0.005$ & $0.054$ & $-0.031$ & $0.049$ \\
25 & --- & --- & --- & --- & --- & --- & $-0.001$ & $0.034$ & $\phantom{-}0.017$ & $0.037$ & $-0.015$ & $0.048$ & $\phantom{-}0.009$ & $0.045$ & $\phantom{-}0.005$ & $0.063$ & $-0.017$ & $0.028$ \\
26 & --- & --- & --- & --- & --- & --- & $\phantom{-}0.001$ & $0.055$ & $\phantom{-}0.019$ & $0.041$ & $-0.008$ & $0.029$ & $\phantom{-}0.009$ & $0.038$ & $\phantom{-}0.007$ & $0.068$ & $-0.035$ & $0.057$ \\
27 & --- & --- & --- & --- & --- & --- & $-0.004$ & $0.072$ & $\phantom{-}0.019$ & $0.040$ & $-0.019$ & $0.060$ & $\phantom{-}0.007$ & $0.035$ & $\phantom{-}0.002$ & $0.034$ & $-0.051$ & $0.081$ \\
28 & --- & --- & --- & --- & --- & --- & $-0.008$ & $0.049$ & $\phantom{-}0.043$ & $0.090$ & $-0.014$ & $0.044$ & $\phantom{-}0.020$ & $0.080$ & $\phantom{-}0.003$ & $0.045$ & $-0.049$ & $0.079$ \\
29 & --- & --- & --- & --- & --- & --- & $-0.001$ & $0.052$ & $\phantom{-}0.029$ & $0.067$ & $-0.008$ & $0.025$ & $\phantom{-}0.012$ & $0.069$ & $\phantom{-}0.005$ & $0.063$ & $-0.047$ & $0.076$ \\
30 & --- & --- & --- & --- & --- & --- & $-0.007$ & $0.068$ & $\phantom{-}0.023$ & $0.051$ & $-0.017$ & $0.054$ & $\phantom{-}0.019$ & $0.076$ & $\phantom{-}0.004$ & $0.067$ & $-0.042$ & $0.068$ \\
31 & --- & --- & --- & --- & --- & --- & --- & --- & --- & --- & --- & --- & $\phantom{-}0.012$ & $0.068$ & $\phantom{-}0.004$ & $0.033$ & $-0.029$ & $0.046$ \\
32 & --- & --- & --- & --- & --- & --- & --- & --- & --- & --- & --- & --- & $\phantom{-}0.010$ & $0.046$ & $\phantom{-}0.004$ & $0.067$ & $-0.022$ & $0.035$ \\
33 & --- & --- & --- & --- & --- & --- & --- & --- & --- & --- & --- & --- & $\phantom{-}0.006$ & $0.035$ & $\phantom{-}0.006$ & $0.074$ & $-0.024$ & $0.040$ \\
34 & --- & --- & --- & --- & --- & --- & --- & --- & --- & --- & --- & --- & $\phantom{-}0.013$ & $0.067$ & $\phantom{-}0.005$ & $0.044$ & $-0.037$ & $0.060$ \\
35 & --- & --- & --- & --- & --- & --- & --- & --- & --- & --- & --- & --- & $\phantom{-}0.011$ & $0.044$ & $\phantom{-}0.005$ & $0.064$ & $-0.044$ & $0.071$ \\
36 & --- & --- & --- & --- & --- & --- & --- & --- & --- & --- & --- & --- & $\phantom{-}0.013$ & $0.066$ & $\phantom{-}0.003$ & $0.035$ & $-0.049$ & $0.079$ \\
37 & --- & --- & --- & --- & --- & --- & --- & --- & --- & --- & --- & --- & $\phantom{-}0.011$ & $0.048$ & $\phantom{-}0.004$ & $0.054$ & $-0.028$ & $0.045$ \\
38 & --- & --- & --- & --- & --- & --- & --- & --- & --- & --- & --- & --- & $\phantom{-}0.018$ & $0.071$ & $\phantom{-}0.003$ & $0.047$ & $-0.051$ & $0.081$ \\
39 & --- & --- & --- & --- & --- & --- & --- & --- & --- & --- & --- & --- & $\phantom{-}0.015$ & $0.068$ & $\phantom{-}0.005$ & $0.065$ & $-0.020$ & $0.032$ \\
40 & --- & --- & --- & --- & --- & --- & --- & --- & --- & --- & --- & --- & $\phantom{-}0.018$ & $0.056$ & $\phantom{-}0.003$ & $0.035$ & $-0.018$ & $0.029$ \\
\bottomrule
\end{tabular}}
\begin{tablenotes}\small
\item Rows correspond to items; columns to simulation conditions. Cells show bias and RMSE for $\alpha_j$. Dashes (---) indicate items outside the admissible change-point window for that condition ($\gamma_j$ is not defined there).
\end{tablenotes}
\end{table}

Table~\ref{tab:sim_secondary_gamma} shows how the near-boundary
phenomenon documented in the primary grid interacts with sample size and test length. Items immediately following the boundary parameter $c$ carry the largest bias and RMSE, with recovery improving steadily as item index increases beyond $c$. The effect of $N$ on near-boundary items is clear for $J = 20$ and $J = 30$. At item 14 ($J = 20$, $c = 12$), bias shrinks
from $-0.141$ at $N = 200$ to essentially zero at $N = 1800$, with RMSE dropping from 0.367 to 0.108. A similar trajectory holds for the
first eligible items in the $J = 30$ conditions, where item 20 goes
from bias $-0.136$ at $N = 200$ to $-0.113$ at $N = 1800$. The convergence is slower, but it is in the same direction.

The $J = 40$ conditions tell a more complex story. Item 26, the
first eligible item under $c = 24$, shows bias of $-0.421$ at $N =
200$ and $-0.184$ at $N = 600$, but $-0.315$ at $N = 1800$, thus worse
than at $N = 600$. This non-monotonicity, which was absent in the shorter-test conditions, likely reflects that $c = 24$ with $J = 40$ leaves only 16 post-change items, and at $\pi = 0.15$ the number of changers contributing information to the near-boundary likelihood is too small to stabilise estimation even at large $N$. Items further from the
boundary in the $J = 40$ conditions recover normally, with bias and
RMSE falling toward zero as both $N$ and item distance from $c$ increase.

\begin{table}[t]
\centering
\caption{Simulation results (secondary grid): change-point effect $\hat{\gamma}_j$. $\pi = 0.15$, $c$ matched to $J$.}
\label{tab:sim_secondary_gamma}
\small
\resizebox{\textwidth}{!}{%
\begin{tabular}{@{}rrrrrrrrrrrrrrrrrrr@{}}
\toprule
Item & \multicolumn{2}{c}{$N=200$, $J=20$} & \multicolumn{2}{c}{$N=600$, $J=20$} & \multicolumn{2}{c}{$N=1800$, $J=20$} & \multicolumn{2}{c}{$N=200$, $J=30$} & \multicolumn{2}{c}{$N=600$, $J=30$} & \multicolumn{2}{c}{$N=1800$, $J=30$} & \multicolumn{2}{c}{$N=200$, $J=40$} & \multicolumn{2}{c}{$N=600$, $J=40$} & \multicolumn{2}{c}{$N=1800$, $J=40$} \\
\cmidrule(lr){2-3}\cmidrule(lr){4-5}\cmidrule(lr){6-7}\cmidrule(lr){8-9}\cmidrule(lr){10-11}\cmidrule(lr){12-13}\cmidrule(lr){14-15}\cmidrule(lr){16-17}\cmidrule(lr){18-19}
 & Bias & RMSE & Bias & RMSE & Bias & RMSE & Bias & RMSE & Bias & RMSE & Bias & RMSE & Bias & RMSE & Bias & RMSE & Bias & RMSE \\
\midrule
1 & --- & --- & --- & --- & --- & --- & --- & --- & --- & --- & --- & --- & --- & --- & --- & --- & --- & --- \\
2 & --- & --- & --- & --- & --- & --- & --- & --- & --- & --- & --- & --- & --- & --- & --- & --- & --- & --- \\
3 & --- & --- & --- & --- & --- & --- & --- & --- & --- & --- & --- & --- & --- & --- & --- & --- & --- & --- \\
4 & --- & --- & --- & --- & --- & --- & --- & --- & --- & --- & --- & --- & --- & --- & --- & --- & --- & --- \\
5 & --- & --- & --- & --- & --- & --- & --- & --- & --- & --- & --- & --- & --- & --- & --- & --- & --- & --- \\
6 & --- & --- & --- & --- & --- & --- & --- & --- & --- & --- & --- & --- & --- & --- & --- & --- & --- & --- \\
7 & --- & --- & --- & --- & --- & --- & --- & --- & --- & --- & --- & --- & --- & --- & --- & --- & --- & --- \\
8 & --- & --- & --- & --- & --- & --- & --- & --- & --- & --- & --- & --- & --- & --- & --- & --- & --- & --- \\
9 & --- & --- & --- & --- & --- & --- & --- & --- & --- & --- & --- & --- & --- & --- & --- & --- & --- & --- \\
10 & --- & --- & --- & --- & --- & --- & --- & --- & --- & --- & --- & --- & --- & --- & --- & --- & --- & --- \\
11 & --- & --- & --- & --- & --- & --- & --- & --- & --- & --- & --- & --- & --- & --- & --- & --- & --- & --- \\
12 & --- & --- & --- & --- & --- & --- & --- & --- & --- & --- & --- & --- & --- & --- & --- & --- & --- & --- \\
13 & --- & --- & --- & --- & --- & --- & --- & --- & --- & --- & --- & --- & --- & --- & --- & --- & --- & --- \\
14 & $-0.141$ & $0.367$ & $-0.028$ & $0.206$ & $\phantom{-}0.001$ & $0.108$ & --- & --- & --- & --- & --- & --- & --- & --- & --- & --- & --- & --- \\
15 & $-0.111$ & $0.240$ & $\phantom{-}0.002$ & $0.124$ & $-0.003$ & $0.059$ & --- & --- & --- & --- & --- & --- & --- & --- & --- & --- & --- & --- \\
16 & $-0.008$ & $0.160$ & $-0.011$ & $0.077$ & $-0.008$ & $0.046$ & --- & --- & --- & --- & --- & --- & --- & --- & --- & --- & --- & --- \\
17 & $-0.041$ & $0.155$ & $-0.004$ & $0.059$ & $-0.002$ & $0.038$ & --- & --- & --- & --- & --- & --- & --- & --- & --- & --- & --- & --- \\
18 & $-0.026$ & $0.124$ & $-0.003$ & $0.067$ & $-0.002$ & $0.034$ & --- & --- & --- & --- & --- & --- & --- & --- & --- & --- & --- & --- \\
19 & $-0.028$ & $0.096$ & $\phantom{-}0.000$ & $0.055$ & $-0.003$ & $0.036$ & --- & --- & --- & --- & --- & --- & --- & --- & --- & --- & --- & --- \\
20 & $-0.008$ & $0.097$ & $\phantom{-}0.008$ & $0.052$ & $-0.001$ & $0.035$ & $-0.136$ & $0.283$ & $-0.143$ & $0.262$ & $-0.113$ & $0.188$ & --- & --- & --- & --- & --- & --- \\
21 & --- & --- & --- & --- & --- & --- & $-0.210$ & $0.383$ & $-0.043$ & $0.173$ & $\phantom{-}0.005$ & $0.127$ & --- & --- & --- & --- & --- & --- \\
22 & --- & --- & --- & --- & --- & --- & $-0.098$ & $0.274$ & $-0.041$ & $0.148$ & $\phantom{-}0.005$ & $0.092$ & --- & --- & --- & --- & --- & --- \\
23 & --- & --- & --- & --- & --- & --- & $-0.088$ & $0.243$ & $-0.019$ & $0.114$ & $-0.004$ & $0.081$ & --- & --- & --- & --- & --- & --- \\
24 & --- & --- & --- & --- & --- & --- & $-0.067$ & $0.181$ & $-0.023$ & $0.087$ & $-0.010$ & $0.061$ & --- & --- & --- & --- & --- & --- \\
25 & --- & --- & --- & --- & --- & --- & $-0.031$ & $0.137$ & $-0.009$ & $0.073$ & $-0.001$ & $0.044$ & --- & --- & --- & --- & --- & --- \\
26 & --- & --- & --- & --- & --- & --- & $-0.046$ & $0.150$ & $-0.002$ & $0.060$ & $-0.007$ & $0.048$ & $-0.421$ & $0.506$ & $-0.184$ & $0.263$ & $-0.315$ & $0.408$ \\
27 & --- & --- & --- & --- & --- & --- & $-0.030$ & $0.128$ & $-0.006$ & $0.055$ & $-0.002$ & $0.033$ & $-0.151$ & $0.309$ & $-0.229$ & $0.387$ & $-0.055$ & $0.160$ \\
28 & --- & --- & --- & --- & --- & --- & $-0.001$ & $0.099$ & $-0.004$ & $0.051$ & $-0.005$ & $0.026$ & $-0.195$ & $0.361$ & $-0.069$ & $0.205$ & $-0.036$ & $0.160$ \\
29 & --- & --- & --- & --- & --- & --- & $-0.019$ & $0.101$ & $-0.008$ & $0.047$ & $-0.004$ & $0.024$ & $-0.131$ & $0.281$ & $-0.038$ & $0.206$ & $-0.025$ & $0.108$ \\
30 & --- & --- & --- & --- & --- & --- & $-0.015$ & $0.104$ & $\phantom{-}0.010$ & $0.041$ & $\phantom{-}0.000$ & $0.028$ & $-0.066$ & $0.237$ & $-0.055$ & $0.171$ & $-0.034$ & $0.120$ \\
31 & --- & --- & --- & --- & --- & --- & --- & --- & --- & --- & --- & --- & $-0.151$ & $0.343$ & $-0.019$ & $0.125$ & $-0.016$ & $0.077$ \\
32 & --- & --- & --- & --- & --- & --- & --- & --- & --- & --- & --- & --- & $-0.034$ & $0.194$ & $-0.024$ & $0.104$ & $-0.016$ & $0.067$ \\
33 & --- & --- & --- & --- & --- & --- & --- & --- & --- & --- & --- & --- & $-0.035$ & $0.179$ & $-0.019$ & $0.096$ & $-0.017$ & $0.069$ \\
34 & --- & --- & --- & --- & --- & --- & --- & --- & --- & --- & --- & --- & $-0.032$ & $0.182$ & $-0.004$ & $0.085$ & $-0.017$ & $0.059$ \\
35 & --- & --- & --- & --- & --- & --- & --- & --- & --- & --- & --- & --- & $-0.019$ & $0.131$ & $-0.013$ & $0.073$ & $-0.013$ & $0.049$ \\
36 & --- & --- & --- & --- & --- & --- & --- & --- & --- & --- & --- & --- & $-0.043$ & $0.111$ & $\phantom{-}0.005$ & $0.054$ & $-0.006$ & $0.033$ \\
37 & --- & --- & --- & --- & --- & --- & --- & --- & --- & --- & --- & --- & $-0.030$ & $0.104$ & $-0.009$ & $0.053$ & $-0.013$ & $0.031$ \\
38 & --- & --- & --- & --- & --- & --- & --- & --- & --- & --- & --- & --- & $-0.015$ & $0.097$ & $-0.002$ & $0.040$ & $-0.004$ & $0.030$ \\
39 & --- & --- & --- & --- & --- & --- & --- & --- & --- & --- & --- & --- & $-0.016$ & $0.078$ & $-0.001$ & $0.047$ & $-0.004$ & $0.025$ \\
40 & --- & --- & --- & --- & --- & --- & --- & --- & --- & --- & --- & --- & $-0.024$ & $0.093$ & $\phantom{-}0.003$ & $0.049$ & $-0.002$ & $0.022$ \\
\bottomrule
\end{tabular}}
\begin{tablenotes}\small
\item Rows correspond to items; columns to simulation conditions. Cells show bias and RMSE for $\gamma_j$. Dashes (---) indicate items outside the admissible change-point window for that condition ($\gamma_j$ is not defined there).
\end{tablenotes}
\end{table}

Table~\ref{tab:sim_secondary_sigma} reports the estimation performance for $\hat{\sigma}_j$. The findings are consistent with what
was observed in the primary grid: bias fluctuates in sign across items
and conditions, RMSE tracks closely with absolute bias, and neither
$N$ nor $J$ produces any systematic directional effect. The range of
bias values, $[-0.108, 0.111]$, is comparable to the primary grid and
does not narrow much as $N$ increases from 200 to 1800, which reinforces the earlier conclusion that the error in $\hat{\sigma}_j$
is not primarily a finite-sample phenomenon. 

Taken across all parameter blocks, the simulation results paint a
relatively clear picture. The estimator recovers $\hat{\beta}_j$,
$\hat{\alpha}_j$, and the structural parameters $\psi_1$, $\psi_2$,
$\psi_3$ with negligible bias and RMSE that scales well with
$N$. The change-point effects $\hat{\gamma}_j$ are recovered reliably except for items immediately following the boundary, where bias and RMSE increases due to the restricted window of eligible
change-point locations. One possible explanation for this elevated bias is that shorter post-change segments provide weaker information for disentangling latent speed effects from item-specific change effects. The residual standard deviations $\hat{\sigma}_j$ show item-specific bias of modest magnitude with no systematic pattern. Change-point location inference, as measured by
MAE($\hat{\tau}$), is most sensitive to test length and the width of
the change-point window rather than to sample size.
\begin{table}[t]
\centering
\caption{Simulation results (secondary grid): residual standard deviation $\hat{\sigma}_j$. $\pi = 0.15$, $c$ matched to $J$.}
\label{tab:sim_secondary_sigma}
\small
\resizebox{\textwidth}{!}{%
\begin{tabular}{@{}rrrrrrrrrrrrrrrrrrr@{}}
\toprule
Item & \multicolumn{2}{c}{$N=200$, $J=20$} & \multicolumn{2}{c}{$N=600$, $J=20$} & \multicolumn{2}{c}{$N=1800$, $J=20$} & \multicolumn{2}{c}{$N=200$, $J=30$} & \multicolumn{2}{c}{$N=600$, $J=30$} & \multicolumn{2}{c}{$N=1800$, $J=30$} & \multicolumn{2}{c}{$N=200$, $J=40$} & \multicolumn{2}{c}{$N=600$, $J=40$} & \multicolumn{2}{c}{$N=1800$, $J=40$} \\
\cmidrule(lr){2-3}\cmidrule(lr){4-5}\cmidrule(lr){6-7}\cmidrule(lr){8-9}\cmidrule(lr){10-11}\cmidrule(lr){12-13}\cmidrule(lr){14-15}\cmidrule(lr){16-17}\cmidrule(lr){18-19}
 & Bias & RMSE & Bias & RMSE & Bias & RMSE & Bias & RMSE & Bias & RMSE & Bias & RMSE & Bias & RMSE & Bias & RMSE & Bias & RMSE \\
\midrule
1 & $\phantom{-}0.001$ & $0.019$ & $-0.070$ & $0.071$ & $\phantom{-}0.094$ & $0.094$ & $\phantom{-}0.050$ & $0.053$ & $-0.060$ & $0.061$ & $-0.061$ & $0.061$ & $-0.081$ & $0.083$ & $-0.015$ & $0.018$ & $\phantom{-}0.081$ & $0.081$ \\
2 & $-0.058$ & $0.060$ & $-0.024$ & $0.026$ & $\phantom{-}0.069$ & $0.069$ & $\phantom{-}0.088$ & $0.090$ & $\phantom{-}0.034$ & $0.036$ & $-0.018$ & $0.018$ & $-0.036$ & $0.041$ & $-0.085$ & $0.085$ & $\phantom{-}0.087$ & $0.087$ \\
3 & $\phantom{-}0.026$ & $0.031$ & $-0.058$ & $0.059$ & $-0.046$ & $0.047$ & $\phantom{-}0.052$ & $0.054$ & $\phantom{-}0.030$ & $0.032$ & $\phantom{-}0.028$ & $0.028$ & $\phantom{-}0.074$ & $0.076$ & $-0.043$ & $0.045$ & $\phantom{-}0.026$ & $0.027$ \\
4 & $-0.094$ & $0.095$ & $-0.046$ & $0.047$ & $\phantom{-}0.067$ & $0.067$ & $-0.042$ & $0.045$ & $\phantom{-}0.067$ & $0.068$ & $-0.015$ & $0.016$ & $-0.051$ & $0.054$ & $-0.010$ & $0.014$ & $-0.004$ & $0.007$ \\
5 & $-0.025$ & $0.031$ & $-0.016$ & $0.019$ & $\phantom{-}0.042$ & $0.043$ & $-0.082$ & $0.084$ & $-0.049$ & $0.050$ & $\phantom{-}0.022$ & $0.023$ & $\phantom{-}0.010$ & $0.021$ & $\phantom{-}0.070$ & $0.070$ & $\phantom{-}0.015$ & $0.017$ \\
6 & $\phantom{-}0.081$ & $0.083$ & $\phantom{-}0.095$ & $0.096$ & $\phantom{-}0.096$ & $0.096$ & $-0.099$ & $0.101$ & $-0.086$ & $0.086$ & $\phantom{-}0.048$ & $0.048$ & $\phantom{-}0.110$ & $0.112$ & $-0.086$ & $0.086$ & $\phantom{-}0.068$ & $0.068$ \\
7 & $-0.044$ & $0.046$ & $\phantom{-}0.033$ & $0.034$ & $-0.054$ & $0.054$ & $\phantom{-}0.030$ & $0.035$ & $-0.038$ & $0.039$ & $\phantom{-}0.079$ & $0.080$ & $\phantom{-}0.105$ & $0.106$ & $-0.033$ & $0.034$ & $-0.094$ & $0.094$ \\
8 & $\phantom{-}0.082$ & $0.084$ & $-0.098$ & $0.099$ & $-0.015$ & $0.016$ & $\phantom{-}0.034$ & $0.037$ & $-0.007$ & $0.012$ & $-0.078$ & $0.078$ & $-0.080$ & $0.082$ & $\phantom{-}0.041$ & $0.042$ & $\phantom{-}0.079$ & $0.080$ \\
9 & $\phantom{-}0.048$ & $0.052$ & $-0.012$ & $0.015$ & $-0.055$ & $0.055$ & $\phantom{-}0.001$ & $0.017$ & $-0.009$ & $0.013$ & $-0.084$ & $0.084$ & $\phantom{-}0.034$ & $0.038$ & $-0.073$ & $0.074$ & $-0.099$ & $0.099$ \\
10 & $\phantom{-}0.031$ & $0.036$ & $-0.014$ & $0.017$ & $-0.055$ & $0.056$ & $-0.004$ & $0.017$ & $-0.019$ & $0.021$ & $-0.099$ & $0.099$ & $\phantom{-}0.024$ & $0.031$ & $\phantom{-}0.108$ & $0.108$ & $\phantom{-}0.006$ & $0.008$ \\
11 & $-0.042$ & $0.045$ & $\phantom{-}0.012$ & $0.016$ & $-0.014$ & $0.015$ & $-0.002$ & $0.016$ & $-0.051$ & $0.052$ & $\phantom{-}0.089$ & $0.089$ & $-0.019$ & $0.027$ & $-0.056$ & $0.057$ & $\phantom{-}0.085$ & $0.086$ \\
12 & $\phantom{-}0.021$ & $0.026$ & $\phantom{-}0.083$ & $0.083$ & $-0.016$ & $0.017$ & $-0.102$ & $0.104$ & $\phantom{-}0.047$ & $0.048$ & $\phantom{-}0.054$ & $0.054$ & $-0.003$ & $0.020$ & $\phantom{-}0.085$ & $0.085$ & $-0.054$ & $0.055$ \\
13 & $\phantom{-}0.077$ & $0.079$ & $-0.059$ & $0.060$ & $-0.022$ & $0.023$ & $\phantom{-}0.036$ & $0.040$ & $\phantom{-}0.012$ & $0.015$ & $\phantom{-}0.083$ & $0.083$ & $-0.074$ & $0.076$ & $\phantom{-}0.106$ & $0.106$ & $\phantom{-}0.081$ & $0.081$ \\
14 & $-0.016$ & $0.023$ & $\phantom{-}0.099$ & $0.099$ & $-0.067$ & $0.067$ & $-0.004$ & $0.019$ & $\phantom{-}0.066$ & $0.067$ & $\phantom{-}0.023$ & $0.024$ & $\phantom{-}0.022$ & $0.027$ & $\phantom{-}0.111$ & $0.112$ & $-0.095$ & $0.095$ \\
15 & $-0.066$ & $0.068$ & $\phantom{-}0.007$ & $0.012$ & $\phantom{-}0.016$ & $0.017$ & $-0.007$ & $0.020$ & $\phantom{-}0.083$ & $0.083$ & $\phantom{-}0.013$ & $0.014$ & $\phantom{-}0.037$ & $0.041$ & $\phantom{-}0.015$ & $0.018$ & $-0.038$ & $0.038$ \\
16 & $-0.032$ & $0.037$ & $\phantom{-}0.065$ & $0.066$ & $\phantom{-}0.009$ & $0.011$ & $-0.007$ & $0.017$ & $\phantom{-}0.063$ & $0.063$ & $\phantom{-}0.087$ & $0.087$ & $\phantom{-}0.073$ & $0.075$ & $\phantom{-}0.110$ & $0.111$ & $-0.054$ & $0.054$ \\
17 & $-0.032$ & $0.037$ & $\phantom{-}0.046$ & $0.047$ & $\phantom{-}0.025$ & $0.026$ & $\phantom{-}0.082$ & $0.083$ & $\phantom{-}0.047$ & $0.048$ & $-0.020$ & $0.021$ & $\phantom{-}0.009$ & $0.020$ & $\phantom{-}0.009$ & $0.013$ & $\phantom{-}0.044$ & $0.044$ \\
18 & $\phantom{-}0.028$ & $0.033$ & $-0.084$ & $0.085$ & $-0.060$ & $0.060$ & $\phantom{-}0.056$ & $0.058$ & $\phantom{-}0.021$ & $0.023$ & $\phantom{-}0.088$ & $0.088$ & $\phantom{-}0.071$ & $0.073$ & $\phantom{-}0.029$ & $0.031$ & $\phantom{-}0.040$ & $0.041$ \\
19 & $-0.008$ & $0.023$ & $\phantom{-}0.053$ & $0.054$ & $\phantom{-}0.019$ & $0.020$ & $\phantom{-}0.023$ & $0.030$ & $\phantom{-}0.079$ & $0.080$ & $\phantom{-}0.057$ & $0.057$ & $-0.063$ & $0.065$ & $-0.078$ & $0.079$ & $\phantom{-}0.044$ & $0.044$ \\
20 & $\phantom{-}0.055$ & $0.059$ & $\phantom{-}0.087$ & $0.088$ & $\phantom{-}0.031$ & $0.032$ & $-0.093$ & $0.094$ & $\phantom{-}0.007$ & $0.012$ & $\phantom{-}0.052$ & $0.053$ & $\phantom{-}0.040$ & $0.043$ & $\phantom{-}0.016$ & $0.019$ & $-0.095$ & $0.096$ \\
21 & --- & --- & --- & --- & --- & --- & $\phantom{-}0.005$ & $0.022$ & $-0.066$ & $0.066$ & $-0.019$ & $0.019$ & $-0.029$ & $0.034$ & $-0.007$ & $0.012$ & $-0.088$ & $0.088$ \\
22 & --- & --- & --- & --- & --- & --- & $-0.004$ & $0.015$ & $\phantom{-}0.013$ & $0.016$ & $\phantom{-}0.075$ & $0.075$ & $-0.064$ & $0.067$ & $\phantom{-}0.106$ & $0.106$ & $\phantom{-}0.072$ & $0.072$ \\
23 & --- & --- & --- & --- & --- & --- & $\phantom{-}0.020$ & $0.025$ & $\phantom{-}0.082$ & $0.082$ & $-0.002$ & $0.007$ & $-0.003$ & $0.020$ & $\phantom{-}0.093$ & $0.093$ & $\phantom{-}0.092$ & $0.093$ \\
24 & --- & --- & --- & --- & --- & --- & $\phantom{-}0.062$ & $0.065$ & $-0.099$ & $0.100$ & $-0.108$ & $0.108$ & $\phantom{-}0.090$ & $0.091$ & $\phantom{-}0.077$ & $0.078$ & $\phantom{-}0.022$ & $0.022$ \\
25 & --- & --- & --- & --- & --- & --- & $\phantom{-}0.051$ & $0.054$ & $\phantom{-}0.014$ & $0.017$ & $-0.105$ & $0.106$ & $-0.051$ & $0.053$ & $\phantom{-}0.022$ & $0.025$ & $-0.016$ & $0.017$ \\
26 & --- & --- & --- & --- & --- & --- & $-0.079$ & $0.080$ & $-0.080$ & $0.080$ & $\phantom{-}0.059$ & $0.059$ & $-0.062$ & $0.064$ & $-0.055$ & $0.056$ & $-0.078$ & $0.078$ \\
27 & --- & --- & --- & --- & --- & --- & $\phantom{-}0.040$ & $0.044$ & $\phantom{-}0.033$ & $0.034$ & $-0.027$ & $0.028$ & $\phantom{-}0.035$ & $0.039$ & $\phantom{-}0.059$ & $0.060$ & $\phantom{-}0.074$ & $0.074$ \\
28 & --- & --- & --- & --- & --- & --- & $-0.074$ & $0.076$ & $-0.079$ & $0.080$ & $\phantom{-}0.012$ & $0.014$ & $\phantom{-}0.003$ & $0.017$ & $-0.070$ & $0.071$ & $-0.021$ & $0.021$ \\
29 & --- & --- & --- & --- & --- & --- & $-0.015$ & $0.024$ & $\phantom{-}0.062$ & $0.063$ & $\phantom{-}0.083$ & $0.084$ & $\phantom{-}0.005$ & $0.016$ & $-0.041$ & $0.043$ & $\phantom{-}0.052$ & $0.052$ \\
30 & --- & --- & --- & --- & --- & --- & $\phantom{-}0.065$ & $0.070$ & $\phantom{-}0.010$ & $0.016$ & $-0.106$ & $0.106$ & $-0.023$ & $0.029$ & $\phantom{-}0.067$ & $0.068$ & $\phantom{-}0.093$ & $0.093$ \\
31 & --- & --- & --- & --- & --- & --- & --- & --- & --- & --- & --- & --- & $-0.015$ & $0.024$ & $-0.082$ & $0.083$ & $-0.099$ & $0.099$ \\
32 & --- & --- & --- & --- & --- & --- & --- & --- & --- & --- & --- & --- & $\phantom{-}0.031$ & $0.035$ & $\phantom{-}0.010$ & $0.014$ & $-0.045$ & $0.045$ \\
33 & --- & --- & --- & --- & --- & --- & --- & --- & --- & --- & --- & --- & $-0.054$ & $0.057$ & $-0.070$ & $0.070$ & $-0.019$ & $0.020$ \\
34 & --- & --- & --- & --- & --- & --- & --- & --- & --- & --- & --- & --- & $\phantom{-}0.088$ & $0.089$ & $-0.030$ & $0.031$ & $\phantom{-}0.057$ & $0.057$ \\
35 & --- & --- & --- & --- & --- & --- & --- & --- & --- & --- & --- & --- & $-0.064$ & $0.066$ & $-0.034$ & $0.036$ & $\phantom{-}0.058$ & $0.058$ \\
36 & --- & --- & --- & --- & --- & --- & --- & --- & --- & --- & --- & --- & $-0.066$ & $0.069$ & $-0.006$ & $0.011$ & $\phantom{-}0.085$ & $0.085$ \\
37 & --- & --- & --- & --- & --- & --- & --- & --- & --- & --- & --- & --- & $-0.033$ & $0.038$ & $-0.022$ & $0.025$ & $\phantom{-}0.074$ & $0.074$ \\
38 & --- & --- & --- & --- & --- & --- & --- & --- & --- & --- & --- & --- & $\phantom{-}0.037$ & $0.042$ & $-0.069$ & $0.070$ & $\phantom{-}0.044$ & $0.045$ \\
39 & --- & --- & --- & --- & --- & --- & --- & --- & --- & --- & --- & --- & $\phantom{-}0.049$ & $0.053$ & $\phantom{-}0.065$ & $0.066$ & $-0.074$ & $0.074$ \\
40 & --- & --- & --- & --- & --- & --- & --- & --- & --- & --- & --- & --- & $\phantom{-}0.023$ & $0.032$ & $\phantom{-}0.008$ & $0.015$ & $-0.081$ & $0.081$ \\
\bottomrule
\end{tabular}}
\begin{tablenotes}\small
\item Rows correspond to items; columns to simulation conditions. Cells show bias and RMSE for $\sigma_j$. Dashes (---) indicate items outside the admissible change-point window for that condition ($\gamma_j$ is not defined there).
\end{tablenotes}
\end{table}


\section{Empirical analysis}\label{sec:empirical}

\subsection{Data and Implementation}

The data come from the Amsterdam Chess Test \citep[ACT;][]{van2005psychometric, LNIRT},
a computer-administered test of tactical chess ability. We analysed the response times for the first $J = 20$ tactical items, recorded in milliseconds and log-transformed. The analysis was based on $N = 256$ respondents\footnote{The code for the empirical analysis can be found here: \url{https://osf.io/86hfx/overview?view_only=53d17590af5e49679068990fdae8a4a9}.}. 

For numerical stability, we used a two-stage optimisation procedure. First, a weakly penalised marginal likelihood was maximised, with ridge penalties on the post-change effects $\gamma_j$ and the change-point location parameter $\psi_1$. The resulting estimates were then used as starting values for an unpenalised maximum likelihood refit. All reported estimates are based on the unpenalised fit.

Since the substantive aim is to identify respondents with interpretable changes in response-time behaviour, we used ICL as the model selection criterion. This criterion was appropriate because it penalises models that improve marginal fit at the cost of highly uncertain respondent-level change-point classifications. Given the moderate sample size and the short response sequence, such a penalty provides a conservative safeguard.

To summarise uncertainty in the respondent-specific change-point locations, we computed the posterior probability of change,
\[
P(\tau_i<J\mid \boldsymbol y_i,\hat{\boldsymbol\theta})
=
1-
P(\tau_i=J\mid \boldsymbol y_i,\hat{\boldsymbol\theta}),
\]
and the normalised posterior entropy
\[
H_i
=
-
\frac{1}{\log(J-c)}
\sum_{\tau=c+1}^{J}
P(\tau_i=\tau\mid \boldsymbol y_i,\hat{\boldsymbol\theta})
\log
P(\tau_i=\tau\mid \boldsymbol y_i,\hat{\boldsymbol\theta}).
\]
The normalisation places $H_i$ on the interval $[0,1]$. Values close to zero indicate that the posterior distribution is concentrated on a small number of possible change-point locations, whereas values close to one indicate greater uncertainty across the admissible locations.

\subsection{Results}


\begin{table}[t]
\centering
\caption{Item parameter estimates with standard errors for the
         change-point response time model. Standard errors in parentheses. All parameters are on the log response time scale. Items RT1--RT6 precede the earliest
         admissible change-point and carry no $\gamma_j$ parameter (---). Significance stars on $\hat{\gamma}_j$ reflect one-sided Wald tests ($H_1\colon \gamma_j < 0$) with Holm correction across the 14 items: $^{**}p < .01$, $^{***}p < .001$.}
\label{tab:item_params}
\small
\begin{tabular}{lcccc}
\toprule
Item & $\hat{\beta}_j$ & $\hat{\alpha}_j$ & $\hat{\gamma}_j$ & $\hat{\sigma}_j$ \\
\midrule
RT1  & 1.543 (0.033) & 0.417 (0.028) & ---                       & 0.324 (0.017) \\
RT2  & 1.366 (0.038) & 0.461 (0.034) & ---                       & 0.396 (0.021) \\
RT3  & 1.857 (0.045) & 0.488 (0.041) & ---                       & 0.528 (0.026) \\
RT4  & 1.958 (0.035) & 0.438 (0.030) & ---                       & 0.355 (0.018) \\
RT5  & 1.654 (0.036) & 0.444 (0.031) & ---                       & 0.368 (0.019) \\
RT6  & 2.063 (0.040) & 0.383 (0.037) & ---                       & 0.506 (0.024) \\
\midrule
RT7  & 2.414 (0.044) & 0.528 (0.040) & $-$0.974 (0.262)$^{***}$ & 0.462 (0.024) \\
RT8  & 2.480 (0.043) & 0.448 (0.039) & $-$0.547 (0.216)$^{**}$  & 0.506 (0.024) \\
RT9  & 2.736 (0.043) & 0.398 (0.041) & $-$0.963 (0.282)$^{***}$ & 0.537 (0.027) \\
RT10 & 2.446 (0.035) & 0.351 (0.031) & $-$0.625 (0.132)$^{***}$ & 0.405 (0.020) \\
RT11 & 2.816 (0.031) & 0.184 (0.030) & $-$0.694 (0.119)$^{***}$ & 0.430 (0.020) \\
RT12 & 2.961 (0.037) & 0.226 (0.036) & $-$0.667 (0.141)$^{***}$ & 0.527 (0.024) \\
RT13 & 2.933 (0.030) & 0.142 (0.029) & $-$0.897 (0.108)$^{***}$ & 0.413 (0.020) \\
RT14 & 2.897 (0.029) & 0.237 (0.028) & $-$0.615 (0.092)$^{***}$ & 0.381 (0.018) \\
RT15 & 2.975 (0.031) & 0.159 (0.030) & $-$0.551 (0.097)$^{***}$ & 0.442 (0.020) \\
RT16 & 3.017 (0.027) & 0.098 (0.027) & $-$0.632 (0.100)$^{***}$ & 0.395 (0.019) \\
RT17 & 2.906 (0.030) & 0.034 (0.029) & $-$0.490 (0.107)$^{***}$ & 0.441 (0.020) \\
RT18 & 2.660 (0.038) & 0.104 (0.036) & $-$0.638 (0.112)$^{***}$ & 0.537 (0.025) \\
RT19 & 3.180 (0.020) & 0.076 (0.019) & $-$0.877 (0.054)$^{***}$ & 0.257 (0.014) \\
RT20 & 2.777 (0.040) & 0.097 (0.035) & $-$0.564 (0.120)$^{***}$ & 0.520 (0.026) \\
\bottomrule
\end{tabular}
\end{table}

Model selection via ICL identified $c = 5$ as the optimal boundary parameter, meaning that the admissible change-point locations are $\tau_i \in \{6, \ldots, 20\}$. Estimated item parameters are presented in Table~\ref{tab:item_params}. The loading estimates $\hat{\alpha}_j$ are positive for all items and decline markedly from RT10 onwards, indicating that individual differences in latent speed account for progressively less of the response time variance in the latter part of the test. The change-point effects $\hat{\gamma}_j$ are negative for all 14 eligible items (RT7--RT20), consistent with a systematic reduction in response times after the inferred change-point. The largest decreases are observed at RT7 ($\hat{\gamma} = -0.974$), RT9 ($\hat{\gamma} = -0.963$), RT13 ($\hat{\gamma} = -0.897$), and RT19 ($\hat{\gamma} = -0.877$). Table~\ref{tab:item_params} also reports one-sided Wald tests for each $\gamma_j$ under $H_1\colon \gamma_j < 0$, with Holm correction across the 14 items. All 14 estimates are negative and statistically significant at the 5\% level. We focus the item-level hypothesis tests on $\gamma_j$, since these parameters directly quantify the post-change shift in response time.

The estimated structural parameters that govern the change-point distribution are reported in Table~\ref{tab:cp_params}. The parameter $\hat{\psi}_1 = 0.009$ is negligible and not statistically different from zero under a likelihood ratio test ($\chi^2(1) = 0.029$, $p = .864$), indicating that change-points are approximately uniformly distributed across admissible locations conditional on changing. The intercept $\hat{\psi}_2 = 1.597$ implies a baseline probability of no change-point of $\widehat{P}(\tau_i = 20 \mid \xi_i = 0) = 0.832$, with a 95\% confidence interval of $[0.774, 0.889]$. The speed-slope parameter $\hat{\psi}_3 = -0.061$ is also not significant ($\chi^2(1) = 0.099$, $p = .753$), indicating that latent speed does not predict whether a respondent exhibits a change-point.

\begin{table}[t]
\centering
\begin{threeparttable}
\caption{Estimates and inference for change-point structural parameters.}
\label{tab:cp_params}
\small
\begin{tabular}{llrrrrrr}
\toprule
 & & \multicolumn{4}{c}{Wald} & \multicolumn{2}{c}{LRT vs.\ $H_0\colon \psi = 0$} \\
\cmidrule(lr){3-6}\cmidrule(lr){7-8}
Parameter & Interpretation & Est. & SE & $z$ & $p$ & $\chi^2(1)$ & $p$ \\
\midrule
$\psi_1$ & CP-location       & $\phantom{-}0.009$ & $0.058$ & $\phantom{-}0.152$ & $.879$ & $0.029$ & $.864$ \\
$\psi_2$ & No-CP log-odds         & $\phantom{-}1.597$ & $0.209$ & $\phantom{-}7.629$ & $<.001$ & \multicolumn{2}{c}{---\tnote{a}} \\
$\psi_3$ & Speed $\times$ no-CP   & $-0.061$           & $0.202$ & $-0.301$           & $.764$  & $0.099$ & $.753$ \\
\midrule
\multicolumn{8}{l}{%
  $\widehat{P}(\text{no CP} \mid \xi = 0) = 0.832$,\quad
  95\% CI $[0.774,\ 0.889]$ (delta method)%
}\\
\bottomrule
\end{tabular}
\begin{tablenotes}
\small
\item[a] The null $\psi_2 = 0$ corresponds to $P(\text{no CP}) = 0.50$ and is not a hypothesis of scientific interest in this study; the estimate and SE are reported instead.
\end{tablenotes}
\end{threeparttable}
\end{table}

Posterior summaries of respondent-level change-points are reported in Tables~\ref{tab:act_tau_summary} and \ref{tab:act_tau_distribution}. The posterior probability of a change-point was highly skewed across respondents: the mean posterior probability of change was 0.169, whereas the median was 0.012. Thus, most respondents had little posterior support for a change-point, while a smaller subset showed strong evidence of a shift in response behaviour. Using the threshold $P(\tau_i < J \mid \mathbf{y}_i) \geq 0.5$, 39 respondents are classified as having changed and 217 as showing no change; 34 respondents exceeded the threshold of $0.8$. Among respondents classified as changed, the average posterior mean change-point was 12.40, suggesting that the shift typically occurred around items 12--13. The mean normalised posterior entropy was 0.306, with a median of 0.070, indicating that posterior uncertainty about $\tau_i$ was generally low but larger for a subset of respondents. The modal posterior estimate equals $J = 20$ (no change) for 220 of 256 respondents (85.9\%). Among the classified changers, the modal change-point locations are spread across the test, with the highest single-location frequencies at $\tau_i = 6$ (6 respondents), $\tau_i = 9$ (5 respondents), and $\tau_i = 13$ (5 respondents); no respondent has a modal estimate at locations 12, 15, or 19.

\begin{table}[t]
\centering
\caption{Posterior summary of respondent-specific change-point inference. Respondents are classified as changed if $P(\tau_i < J \mid \mathbf{y}_i) \geq 0.5$.}
\label{tab:act_tau_summary}
\small
\begin{tabular}{lr}
\toprule
Metric & Value \\
\midrule
Sample size $N$                                       & 256    \\
Number of items $J$                                   & 20     \\
Selected $c$                                          & 5      \\
Earliest admissible change-point                      & item 6 \\
\midrule
Mean $P(\tau_i<J\mid \mathbf{y}_i)$                   & 0.169  \\
Median $P(\tau_i<J\mid \mathbf{y}_i)$                 & 0.012  \\
Respondents with $P(\tau_i<J\mid \mathbf{y}_i)\geq .50$ & 39   \\
Respondents with $P(\tau_i<J\mid \mathbf{y}_i)\geq .80$ & 34   \\
\midrule
Proportion with modal $\hat{\tau}_i=J$ (no change)    & 0.859  \\
Mean posterior mean $E[\tau_i\mid \mathbf{y}_i]$      & 18.76  \\
Mean posterior mean among classified changers         & 12.40  \\
\midrule
Mean normalised posterior entropy                     & 0.306  \\
Median normalised posterior entropy                   & 0.070  \\
\midrule
Classified as changed                                 & 39     \\
Classified as no change                               & 217    \\
\bottomrule
\end{tabular}
\end{table}

\begin{table}[t]
\centering
\caption{Distribution of modal posterior change-point
         $\hat{\tau}_i = \arg\max_\tau\, P(\tau_i = \tau \mid \mathbf{y}_i)$.
         $\hat{\tau}_i = 20$ indicates the no-change state.
         Locations 12, 15, and 19 are not the modal estimate for any respondent.}
\label{tab:act_tau_distribution}
\small
\begin{tabular}{rr}
\toprule
Modal $\hat{\tau}_i$ & Frequency \\
\midrule
 6  &   6 \\
 7  &   3 \\
 8  &   1 \\
 9  &   5 \\
10  &   3 \\
11  &   1 \\
13  &   5 \\
14  &   2 \\
16  &   2 \\
17  &   3 \\
18  &   5 \\
20  & 220 \\
\midrule
Total & 256 \\
\bottomrule
\end{tabular}
\end{table}

\begin{table}[t]
\centering
\caption{Model-implied prior and average posterior probabilities for
         change-point locations. The prior is evaluated at $\xi_i = 0$
         and is determined by $\hat{\psi}_1$, $\hat{\psi}_2$, $\hat{\psi}_3$.
         The average posterior is
         $N^{-1}\sum_i P(\tau_i = \tau \mid \mathbf{y}_i)$.}
\label{tab:act_prior_posterior_tau}
\small
\begin{tabular}{rrr}
\toprule
$\tau$ & Prior & Avg.\ posterior \\
\midrule
 6  & 0.0113 & 0.0194 \\
 7  & 0.0114 & 0.0106 \\
 8  & 0.0115 & 0.0099 \\
 9  & 0.0117 & 0.0158 \\
10  & 0.0118 & 0.0117 \\
11  & 0.0119 & 0.0075 \\
12  & 0.0120 & 0.0060 \\
13  & 0.0121 & 0.0107 \\
14  & 0.0122 & 0.0077 \\
15  & 0.0123 & 0.0064 \\
16  & 0.0124 & 0.0136 \\
17  & 0.0125 & 0.0135 \\
18  & 0.0126 & 0.0210 \\
19  & 0.0127 & 0.0147 \\
20  & 0.8317 & 0.8314 \\
\midrule
Total & 1.0000 & 1.0000 \\
\bottomrule
\end{tabular}
\end{table}

\begin{figure}[htbp]
    \centering
    \includegraphics[width=.75\textwidth]{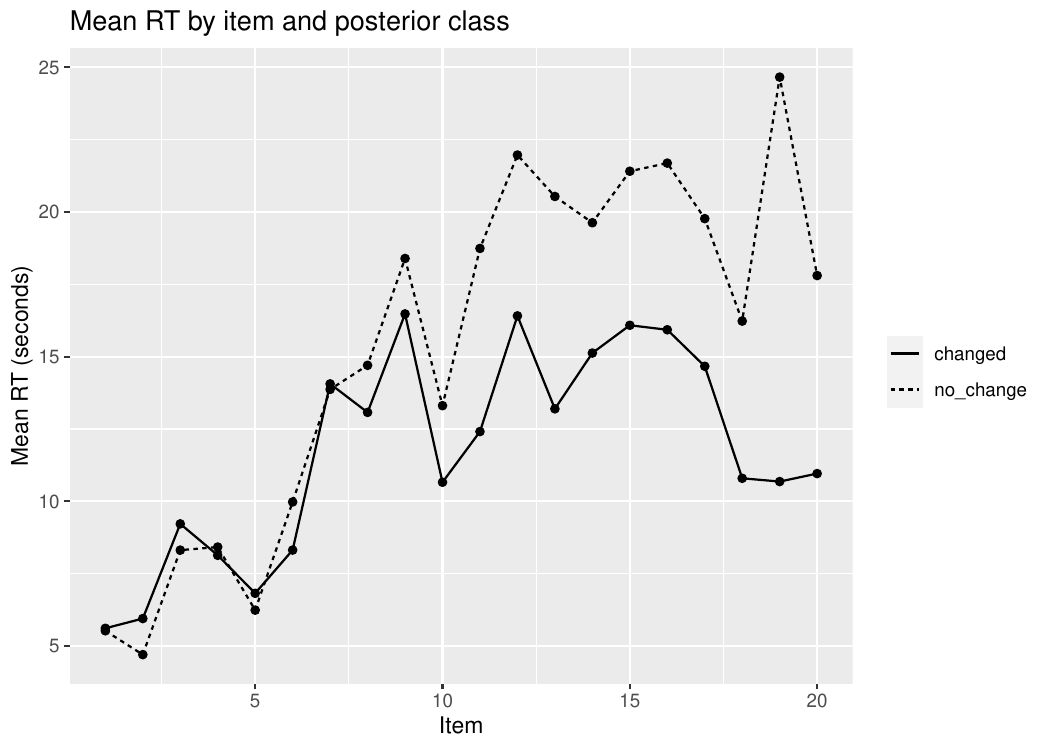}
    \caption{RTs for unchanged and changed respondents.}
    \label{fig:rt-by-class}
\end{figure}

Figure~\ref{fig:rt-by-class} plots mean item response times separately for the two groups. The changed and unchanged groups are closely matched on items RT1--RT6, consistent with the absence of a change-point effect in that window. From RT7 onwards, the changed group responds consistently and substantially faster: the gap grows from approximately 1--2 seconds on early items to 6--9 seconds on items RT11--RT16, and the no-change group continues to slow down through RT19 (mean 24.7 seconds) while the changed group plateaus around 10--15 seconds. 

\begin{figure}[htbp]
    \centering
    \includegraphics[width=.75\textwidth]{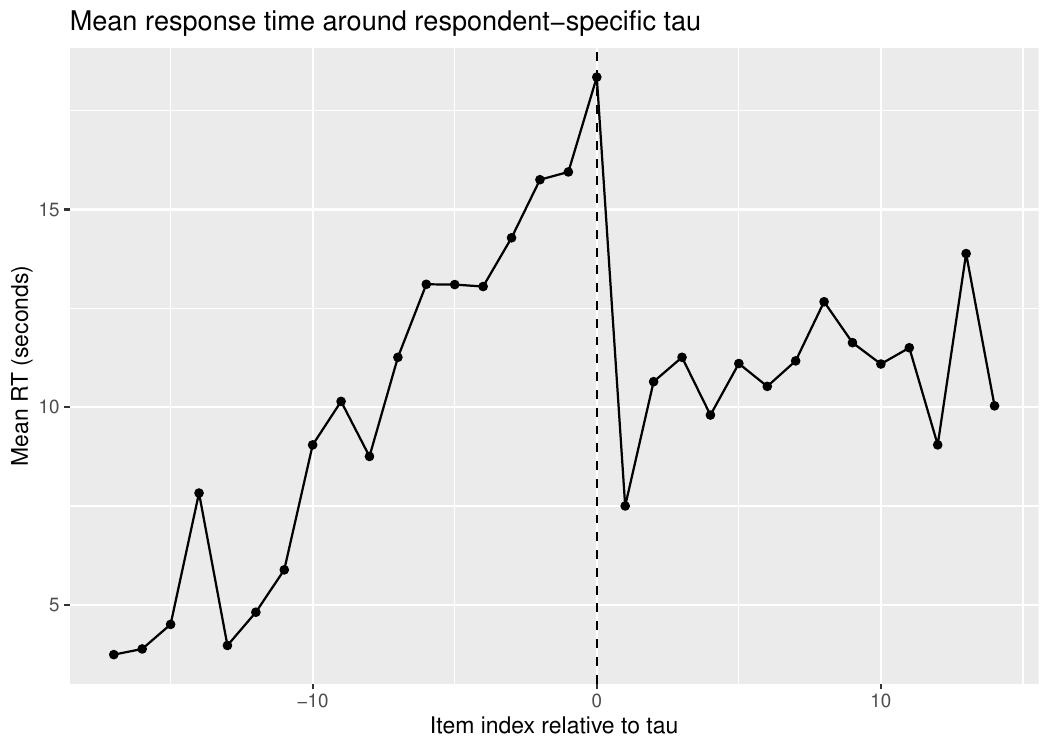}
    \caption{RTs around $\tau$.}
    \label{fig:p3-rt-tau}
\end{figure}

Figure~\ref{fig:p3-rt-tau} aligns each respondent to their own estimated change-point. Mean response times in the changed group are stable at approximately 13--15 seconds in the items preceding $\hat{\tau}_i$, then fall sharply at the estimated change-point and stabilise at approximately 10--11 seconds thereafter. The abrupt transition is consistent with the discrete change-point formulation of the model and argues against a gradual drift in response speed across the test.

\begin{figure}[htbp]
    \centering
    \includegraphics[width=.75\textwidth]{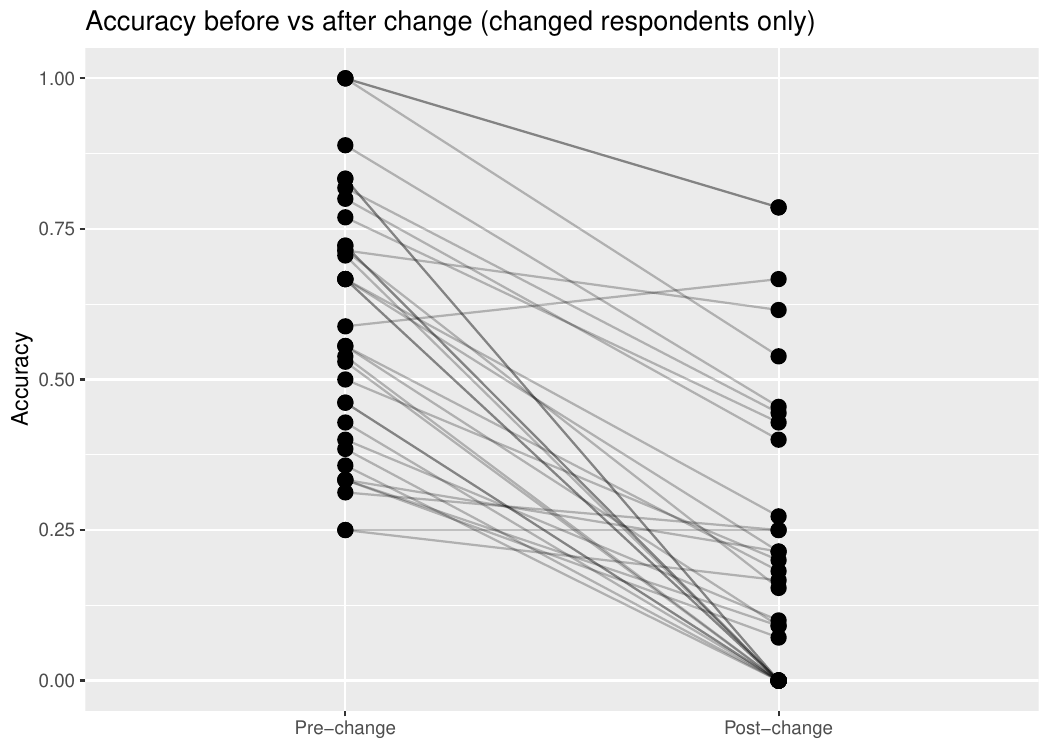}
    \caption{Mean accuracy before and after the estimated change-point for respondents classified as changed. Each line connects one respondent's pre- and post-change accuracy, based on that respondent's modal estimated change-point.}
    \label{fig:p2-acc}
\end{figure}

A paired $t$-test compared mean item accuracy before and after the estimated change-point among the 39 classified changers, using each respondent's modal posterior $\hat{\tau}_i$ to define the two windows. Pre-change accuracy was significantly higher than post-change accuracy ($t(35) = 10.53$, $p < .001$), with a mean within-person decrease of 0.40 (95\% CI $[0.32, 0.48]$). Figure~\ref{fig:p2-acc} illustrates this decline: everyone except for one individual has a downward facing trajectory. The classified changers thus shift to a faster and less accurate response strategy at their estimated change-point.

\begin{figure}
    \centering
    \includegraphics[width=0.75\linewidth]{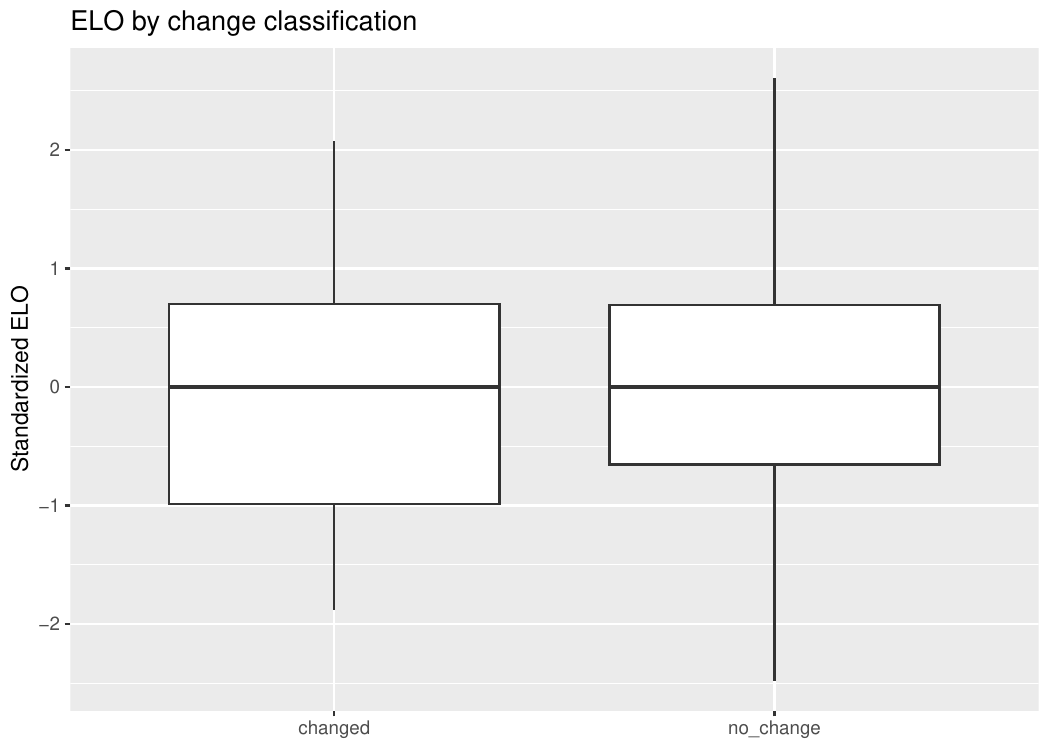}
    \caption{Boxplots of ELO score for the changed and unchanged respondents.}
    \label{fig:boxplot_ELO}
\end{figure}

Lastly, Figure~\ref{fig:boxplot_ELO} examines whether change-point behaviour is associated with chess ability as measured by the standardised ELO rating. The two groups are very similar: both have a median ELO of zero, means of $-0.07$ and $0.01$ respectively, and relatively similar interquartile ranges, although the changed group has less ELO variation. A two-sample $t$-test finds no significant difference ($t(50.3) = -0.43$, $p = .670$), indicating that the tendency to shift response strategy is unrelated to chess ability.

\section{Discussion}\label{sec:discussion}

This paper proposed a latent variable model for log response times with individual-specific change-points. The model extends the log-normal response time model by adding an item-specific shift after an unobserved change-point. The simulation study shows that the proposed model recovers item parameters and structural parameters with good accuracy across a range of conditions. Recovery of the change-point locations depends on the width of the admissible region and the prevalence of respondents exhibiting a change. When the range of possible change-point locations is narrow, individual change-points are estimated with high precision. When the range is wider, estimation becomes more difficult, particularly for items close to the lower boundary of the admissible region. This is reflected in the recovery of the change-point effect parameters, where bias is largest for items immediately following the boundary and decreases for items further away. 

An important aspect of the proposed approach is that inference on change-point locations is based on a full posterior distribution.  This contrasts with procedures that produce a single detected point without an accompanying measure of uncertainty. The likelihood-based formulation also supports statistical inference for the parameters governing both the response time model and the change-point process. Standard errors, confidence intervals and hypothesis tests can be obtained for the change-point effect parameters and for the parameters controlling the distribution of change-points. This enables assessment of the magnitude of changes in response behaviour and the factors associated with their occurrence. In this respect, the proposed framework differs from approaches that focus on detection alone and do not provide parameter-level inference.

The model does not identify the psychological cause of a change-point. A negative post-change effect may reflect rapid guessing, a strategic shift, reduced engagement, or external time pressure. The empirical analysis illustrates this point: the estimated changers responded faster after the change-point and also showed lower post-change accuracy, but the model itself only estimates the timing and magnitude of the response-time shift. This allows the framework to be applied across different types of measurement instruments and testing conditions.  

Another feature of the proposed framework is that it is largely agnostic to the form of the observed response outcomes. Because the model is specified for the response time process and the latent change-point structure, it can in principle be combined with a wide range of measurement models for response accuracy, including binary, ordinal, nominal, or multidimensional item response models. In contrast, many existing change-point approaches are developed for specific outcome types. Hence, the proposed modeling framework provides a method for studying within-person changes in response behaviour in a broad class of psychometric settings.

This framework may also be useful for data quality assessment. Abrupt changes in response times may reflect disengagement, rapid guessing, careless responding, or shifts in test-taking strategy. By estimating individual-specific change-points, the model provides a way to identify respondents whose response processes deviate from a baseline model over the course of an assessment. Such information could be incorporated into person-fit analyses, validity investigations, or procedures for detecting aberrant responding in large-scale assessments and survey data.

There are limitations to the proposed framework. First, the model assumes a single change-point for each respondent. In settings where multiple changes occur, a more general formulation could be considered. Second, the change-point is restricted to a discrete set of item indices and depends on the ordering of items. Extensions to more flexible or continuous representations of change may be of interest. The proposed framework can be extended in several directions. One natural extension is to allow multiple change-points within a hierarchical or hidden-state formulation. Another direction is to integrate response accuracy and response times within a joint model, as in \cite{zhang2026joint, lu2026detecting}, that allows for changes in both processes. Further extensions may incorporate covariates at the person or item level to explain variability in the probability and location of change-points. From a computational perspective, alternative estimation approaches such as stochastic approximation \citep{alfonzetti2025composite} or variational methods \citep{lyu2025multi} may be useful in large-scale applications, as has been proposed in the differential item functioning literature which is closely related.

In summary, the model treats changes in response speed as respondent-specific latent events rather than as fixed detected points. This makes it possible to estimate item-level post-change effects, the distribution of change-point locations, and respondent-level posterior probabilities within the same likelihood-based framework. In the empirical application, this distinguished a small group of respondents whose response times decreased sharply after an estimated change-point, while retaining uncertainty about the timing and presence of that change.

\clearpage

\bibliography{bibliography}

@article{van2005psychometric,
  title={A psychometric analysis of chess expertise},
  author={Van Der Maas, Han LJ and Wagenmakers, Eric-Jan},
  journal={The American journal of psychology},
  volume={118},
  number={1},
  pages={29--60},
  year={2005},
  publisher={University of Illinois Press}
}

@article{wallin2025latent,
  title={A latent variable model with change-points and its application to time pressure effects in educational assessment},
  author={Wallin, Gabriel and Chen, Yunxiao and Lee, Yi-Hsuan and Li, Xiaoou},
  journal={The Annals of Applied Statistics},
  volume={19},
  number={3},
  pages={2490--2516},
  year={2025},
  publisher={Institute of Mathematical Statistics}
}

@article{van2008using,
  title={Using response times for item selection in adaptive testing},
  author={van der Linden, Wim J},
  journal={Journal of Educational and Behavioral Statistics},
  volume={33},
  number={1},
  pages={5--20},
  year={2008}
}

@article{lee2014modeling,
  title={Modeling response times in item response theory},
  author={Lee, Yi-Hsuan and Jia, Yan},
  journal={Educational and Psychological Measurement},
  volume={74},
  number={2},
  pages={321--340},
  year={2014}
}

@article{fan2012does,
  title={Does response time matter? A study of its effects on test performance},
  author={Fan, Jian and Wang, Wei-Cheng and others},
  journal={Educational Measurement: Issues and Practice},
  volume={31},
  number={2},
  pages={21--30},
  year={2012}
}

@article{wise2006modeling,
  title={An application of item response time: The effort-moderated IRT model},
  author={Wise, Steven L and DeMars, Christine E},
  journal={Journal of Educational Measurement},
  volume={43},
  number={1},
  pages={19--38},
  year={2006}
}

@article{wise2010rapid,
  title={Examinee noneffort and the validity of program assessment results},
  author={Wise, Steven L and DeMars, Christine E},
  journal={Educational Assessment},
  volume={15},
  number={1},
  pages={27--41},
  year={2010}
}

@article{wise2012detecting,
  title={Detecting rapid-guessing behavior in computer-based tests},
  author={Wise, Steven L and Ma, Lina},
  journal={Applied Measurement in Education},
  volume={25},
  number={3},
  pages={241--257},
  year={2012}
}

@article{vanderguo2008,
  title={A hierarchical model for response times in tests},
  author={van der Linden, Wim J and Guo, Feng},
  journal={Psychometrika},
  volume={73},
  number={2},
  pages={181--199},
  year={2008}
}

@article{marianti2014modeling,
  title={Modeling response times and accuracy in tests with rapid guessing behavior},
  author={Marianti, Siska and Fox, Jean-Paul and Avetisyan, Mariam},
  journal={British Journal of Mathematical and Statistical Psychology},
  volume={67},
  number={2},
  pages={276--299},
  year={2014}
}

@article{sinharay2017detecting,
  title={Detecting test speededness using response times},
  author={Sinharay, Sandip},
  journal={Applied Psychological Measurement},
  volume={41},
  number={2},
  pages={121--138},
  year={2017}
}

@article{shao2016detecting,
  title={Detecting test speededness using change-point analysis},
  author={Shao, Chenguang and others},
  journal={Journal of Educational Measurement},
  volume={53},
  number={4},
  pages={411--429},
  year={2016}
}

@article{zhu2023bayesian,
  title={Bayesian change-point analysis approach to detecting aberrant test-taking behavior using response times},
  author={Zhu, Hongyue and Jiao, Hong and Gao, Wei and Meng, Xiangbin},
  journal={Journal of Educational and Behavioral Statistics},
  volume={48},
  number={4},
  pages={490--520},
  year={2023},
  publisher={SAGE Publications Sage CA: Los Angeles, CA}
}

@article{du2025detecting,
  title={Detecting compromised items with response times using a Bayesian change-point approach},
  author={Du, Yang and Zhang, Susu},
  journal={Journal of Educational and Behavioral Statistics},
  volume={50},
  number={2},
  pages={296--330},
  year={2025},
  publisher={Sage Publications Sage CA: Los Angeles, CA}
}

@article{alfonzetti2025composite,
  title={When composite likelihood meets stochastic approximation},
  author={Alfonzetti, Giuseppe and Bellio, Ruggero and Chen, Yunxiao and Moustaki, Irini},
  journal={Journal of the American Statistical Association},
  volume={120},
  number={551},
  pages={1906--1918},
  year={2025},
  publisher={Taylor \& Francis}
}

@article{lyu2025multi,
  title={Multi-group regularized Gaussian variational estimation: Fast detection of DIF},
  author={Lyu, Weicong and Wang, Chun and Xu, Gongjun},
  journal={Psychometrika},
  volume={90},
  number={1},
  pages={2--23},
  year={2025},
  publisher={Cambridge University Press}
}

@article{lu2026detecting,
  title={Detecting Test Speededness Using Responses and/or Response Times: Change Point Analysis Approaches Based on Schwarz Information Criterion},
  author={Lu, Jing and Wang, Chun and Zhang, Jiwei and Liu, Zefeng},
  journal={Psychometrika},
  pages={1--30},
  year={2026},
  publisher={Cambridge University Press}
}

@article{zhang2026joint,
  title={A Joint Model for Graded Responses and Response Times},
  author={Zhang, Xinyu and Meng, Xiangbin and Gao, Wei and Xu, Gongjun},
  journal={Psychometrika},
  pages={1--25},
  year={2026},
  publisher={Cambridge University Press}
}

@article{yu2022cusum,
  title={CUSUM-based methods for detecting test speededness: A comparative study},
  author={Yu, Yi and Cheng, Yanyan},
  journal={Psychometrika},
  volume={87},
  number={3},
  pages={987--1012},
  year={2022}
}

@article{page1954continuous,
  title={Continuous inspection schemes},
  author={Page, Ewan S},
  journal={Biometrika},
  volume={41},
  number={1/2},
  pages={100--115},
  year={1954},
  publisher={JSTOR}
}

@article{cheng2022application,
  title={Application of change point analysis of response time data to detect test speededness},
  author={Cheng, Ying and Shao, Can},
  journal={Educational and Psychological Measurement},
  volume={82},
  number={5},
  pages={1031--1062},
  year={2022},
  publisher={Sage Publications Sage CA: Los Angeles, CA}
}

@article{sinharay2016person,
  title={Person fit analysis in computerized adaptive testing using tests for a change point},
  author={Sinharay, Sandip},
  journal={Journal of Educational and Behavioral Statistics},
  volume={41},
  number={5},
  pages={521--549},
  year={2016},
  publisher={Sage Publications Sage CA: Los Angeles, CA}
}

@article{fox2016joint,
  title={Joint modeling of ability and differential speed using responses and response times},
  author={Fox, Jean-Paul and Marianti, Sukaesi},
  journal={Multivariate behavioral research},
  volume={51},
  number={4},
  pages={540--553},
  year={2016},
  publisher={Taylor \& Francis}
}

@article{van2006lognormal,
  title={A lognormal model for response times on test items},
  author={Van der Linden, Wim J},
  journal={Journal of Educational and Behavioral Statistics},
  volume={31},
  number={2},
  pages={181--204},
  year={2006},
  publisher={Sage Publications Sage CA: Los Angeles, CA}
}

@article{goegebeur2010person,
  title={Person fit for test speededness},
  author={Goegebeur, Yuri and De Boeck, Paul and Molenberghs, Geert},
  journal={Methodology},
  year={2010},
  publisher={Hogrefe Publishing}
}

@article{wang2015mixture,
  title={A mixture hierarchical model for response times and response accuracy},
  author={Wang, Chun and Xu, Gongjun},
  journal={British Journal of Mathematical and Statistical Psychology},
  volume={68},
  number={3},
  pages={456--477},
  year={2015},
  publisher={Wiley Online Library}
}

@article{wise2005response,
  title={Response time effort: A new measure of examinee motivation in computer-based tests},
  author={Wise, Steven L and Kong, Xiaojing},
  journal={Applied Measurement in Education},
  volume={18},
  number={2},
  pages={163--183},
  year={2005},
  publisher={Taylor \& Francis}
}

@article{van1999using,
  title={Using response-time constraints to control for differential speededness in computerized adaptive testing},
  author={Van der Linden, Wim J and Scrams, David J and Schnipke, Deborah L},
  journal={Applied psychological measurement},
  volume={23},
  number={3},
  pages={195--210},
  year={1999},
  publisher={Sage Publications Sage CA: Thousand Oaks, CA}
}

@article{schnipke1997modeling,
  title={Modeling item response times with a two-state mixture model: A new method of measuring speededness},
  author={Schnipke, Deborah L and Scrams, David J},
  journal={Journal of Educational Measurement},
  volume={34},
  number={3},
  pages={213--232},
  year={1997},
  publisher={Wiley Online Library}
}

@Manual{LNIRT,
    title = {LNIRT: LogNormal Response Time Item Response Theory Models},
    author = {Jean-Paul Fox and Konrad Klotzke and Rinke Klein Entink},
    year = {2021},
    note = {R package version 0.5.1},
    url = {https://CRAN.R-project.org/package=LNIRT},
  }

@article{meade2012identifying,
  title={Identifying careless responses in survey data.},
  author={Meade, Adam W and Craig, S Bartholomew},
  journal={Psychological methods},
  volume={17},
  number={3},
  pages={437},
  year={2012},
  publisher={American Psychological Association}
}

\clearpage

\appendix

\section{Appendix: Score Equations}\label{app:score}

The observed-data score can be expressed as a posterior expectation of the complete-data score. For respondent $i$, define the posterior distribution of the latent variables as
\[
p_i(\tau,\xi \mid \boldsymbol{y}_i,\boldsymbol{\theta})
=
\frac{
f(\boldsymbol{y}_i \mid \xi,\tau,\boldsymbol{\theta})
P(\tau \mid \xi,\boldsymbol{\psi})
\phi(\xi)
}{
\sum_{\tau'=c+1}^{J}
\int
f(\boldsymbol{y}_i \mid u,\tau',\boldsymbol{\theta})
P(\tau' \mid u,\boldsymbol{\psi})
\phi(u)
\,du
}.
\]
Then, for a generic parameter $\theta_r$,
\[
\frac{\partial \ell(\boldsymbol{\theta})}{\partial \theta_r}
=
\sum_{i=1}^{N}
\sum_{\tau=c+1}^{J}
\int
p_i(\tau,\xi \mid \boldsymbol{y}_i,\boldsymbol{\theta})
\frac{\partial}{\partial \theta_r}
\log
\left[
f(\boldsymbol{y}_i \mid \xi,\tau,\boldsymbol{\theta})
P(\tau \mid \xi,\boldsymbol{\psi})
\right]
\,d\xi .
\]
Thus, the observed-data score is obtained by averaging the complete-data score over the posterior distribution of the latent speed and change-point location.

The integral over $\xi$ is approximated using Gaussian quadrature. Let $\{\xi^{(k)}\}_{k=1}^{K}$ and $\{w^{(k)}\}_{k=1}^{K}$ denote the quadrature nodes and weights. The corresponding quadrature-based posterior weight for respondent $i$, change-point location $\tau$, and quadrature node $\xi^{(k)}$ is
\[
p_{i\tau k}(\boldsymbol{\theta})
=
\frac{
w^{(k)}
f(\boldsymbol{y}_i \mid \xi^{(k)},\tau,\boldsymbol{\theta})
P(\tau \mid \xi^{(k)},\boldsymbol{\psi})
}{
\sum_{\tau'=c+1}^{J}
\sum_{k'=1}^{K}
w^{(k')}
f(\boldsymbol{y}_i \mid \xi^{(k')},\tau',\boldsymbol{\theta})
P(\tau' \mid \xi^{(k')},\boldsymbol{\psi})
}.
\]
The quadrature-based score equations are then obtained by replacing the posterior integral over $\xi$ by the corresponding weighted sum over quadrature nodes.

Let
\[
r_{ij}(\xi^{(k)},\tau)
=
y_{ij}
-
\beta_j
+
\alpha_j \xi^{(k)}
-
\gamma_j \mathbb{I}(j>\tau)
\]
denote the conditional residual. Under the normal response-time model, the score for the time-intensity parameter $\beta_j$ is
\[
\frac{\partial \ell(\boldsymbol{\theta})}{\partial \beta_j}
=
\sum_{i=1}^{N}
\sum_{\tau=c+1}^{J}
\sum_{k=1}^{K}
p_{i\tau k}(\boldsymbol{\theta})
\frac{r_{ij}(\xi^{(k)},\tau)}{\sigma_j^2}.
\]

The score for the speed-discrimination parameter $\alpha_j$ is
\[
\frac{\partial \ell(\boldsymbol{\theta})}{\partial \alpha_j}
=
-
\sum_{i=1}^{N}
\sum_{\tau=c+1}^{J}
\sum_{k=1}^{K}
p_{i\tau k}(\boldsymbol{\theta})
\frac{\xi^{(k)} r_{ij}(\xi^{(k)},\tau)}{\sigma_j^2}.
\]

For $j>c+1$, the score for the change-point effect parameter $\gamma_j$ is
\[
\frac{\partial \ell(\boldsymbol{\theta})}{\partial \gamma_j}
=
\sum_{i=1}^{N}
\sum_{\tau=c+1}^{j-1}
\sum_{k=1}^{K}
p_{i\tau k}(\boldsymbol{\theta})
\frac{r_{ij}(\xi^{(k)},\tau)}{\sigma_j^2}.
\]
For $j\leq c+1$, $\gamma_j$ is fixed at zero and is not estimated.

The score for the residual standard deviation $\sigma_j$ is
\[
\frac{\partial \ell(\boldsymbol{\theta})}{\partial \sigma_j}
=
\sum_{i=1}^{N}
\sum_{\tau=c+1}^{J}
\sum_{k=1}^{K}
p_{i\tau k}(\boldsymbol{\theta})
\left\{
\frac{r_{ij}(\xi^{(k)},\tau)^2}{\sigma_j^3}
-
\frac{1}{\sigma_j}
\right\}.
\]

It remains to give the score contributions for the parameters governing the change-point distribution. Define
\[
q^{(k)}
=
P(\tau_i=J\mid \xi^{(k)},\boldsymbol{\psi})
=
\frac{1}{1+\exp(-\psi_2-\psi_3\xi^{(k)})},
\]
so that
\[
1-q^{(k)}
=
P(\tau_i<J\mid \xi^{(k)},\boldsymbol{\psi}).
\]
Also define
\[
Z(\psi_1)
=
\sum_{\ell=0}^{J-c-2}\exp(\ell\psi_1),
\]
and, for $\tau\in\{c+1,\ldots,J-1\}$,
\[
a_\tau=\tau-c-1.
\]
Then
\[
P(\tau_i=\tau\mid \xi^{(k)},\boldsymbol{\psi})
=
\frac{\exp(a_\tau\psi_1)}{Z(\psi_1)}
\{1-q^{(k)}\},
\qquad \tau=c+1,\ldots,J-1,
\]
whereas
\[
P(\tau_i=J\mid \xi^{(k)},\boldsymbol{\psi})
=
q^{(k)}.
\]

For $\psi_1$, the derivative is nonzero only when $\tau<J$. Thus,
\[
\frac{\partial}{\partial \psi_1}
\log P(\tau_i=\tau\mid \xi^{(k)},\boldsymbol{\psi})
=
a_\tau
-
\frac{
\sum_{\ell=0}^{J-c-2}\ell\exp(\ell\psi_1)
}{
\sum_{\ell=0}^{J-c-2}\exp(\ell\psi_1)
},
\qquad \tau=c+1,\ldots,J-1,
\]
and
\[
\frac{\partial}{\partial \psi_1}
\log P(\tau_i=J\mid \xi^{(k)},\boldsymbol{\psi})
=
0.
\]

For $\psi_2$ and $\psi_3$, the derivatives depend on whether a change-point occurs. If $\tau<J$, then
\[
\log P(\tau_i=\tau\mid \xi^{(k)},\boldsymbol{\psi})
=
a_\tau\psi_1
-
\log Z(\psi_1)
+
\log\{1-q^{(k)}\}.
\]
Therefore,
\[
\frac{\partial}{\partial \psi_2}
\log P(\tau_i=\tau\mid \xi^{(k)},\boldsymbol{\psi})
=
-q^{(k)},
\qquad \tau=c+1,\ldots,J-1,
\]
and
\[
\frac{\partial}{\partial \psi_3}
\log P(\tau_i=\tau\mid \xi^{(k)},\boldsymbol{\psi})
=
-\xi^{(k)}q^{(k)},
\qquad \tau=c+1,\ldots,J-1.
\]

If $\tau=J$, then
\[
\log P(\tau_i=J\mid \xi^{(k)},\boldsymbol{\psi})
=
\log q^{(k)},
\]
and hence
\[
\frac{\partial}{\partial \psi_2}
\log P(\tau_i=J\mid \xi^{(k)},\boldsymbol{\psi})
=
1-q^{(k)},
\]
and
\[
\frac{\partial}{\partial \psi_3}
\log P(\tau_i=J\mid \xi^{(k)},\boldsymbol{\psi})
=
\xi^{(k)}\{1-q^{(k)}\}.
\]

Combining these complete-data score contributions with the quadrature-based posterior weights gives
\[
\frac{\partial \ell(\boldsymbol{\theta})}{\partial \psi_m}
=
\sum_{i=1}^{N}
\sum_{\tau=c+1}^{J}
\sum_{k=1}^{K}
p_{i\tau k}(\boldsymbol{\theta})
\frac{\partial}{\partial \psi_m}
\log P(\tau\mid \xi^{(k)},\boldsymbol{\psi}),
\qquad m=1,2,3.
\]
\end{document}